\documentclass{article}
\usepackage{spconf,amsmath,epsfig}
\usepackage[colorlinks,
            linkcolor=black,
            anchorcolor=white,
            citecolor=black]{hyperref}
\usepackage{subcaption}
\usepackage{amsmath}
\usepackage{amssymb}

\let\OLDthebibliography\thebibliography
\renewcommand\thebibliography[1]{
  \OLDthebibliography{#1}
  \setlength{\parskip}{0pt}
  \setlength{\itemsep}{0pt plus 0.3ex}
}

\begin{document}\sloppy

\def\x{{\mathbf x}}
\def\L{{\cal L}}

\title{Powerful Lossy Compression for Noisy Images}
%
\name{Shilv Cai\textsuperscript{\rm 1,2}, Xiaoguo Liang\textsuperscript{\rm 1,2}, Shuning Cao\textsuperscript{\rm 1,2},  Luxin Yan\textsuperscript{\rm 1,2}, Sheng Zhong\textsuperscript{\rm 1,2}, Liqun Chen\textsuperscript{\rm 1,2,$\dagger$}\thanks{\textsuperscript{$\dagger$}Corresponding Author.}, Xu Zou\textsuperscript{\rm 1,2}}
\address{\textsuperscript{\rm 1} Huazhong University of Science and Technology, China\\
                    \textsuperscript{\rm 2} National Key Laboratory of Multispectral Information Intelligent Processing Technology, China\\
                    \{caishilv, liangxiaoguo, sn\_cao,  yanluxin, zhongsheng,  chenliqun, zoux\}@hust.edu.cn
                    }
\maketitle
\begin{abstract}
Image compression and denoising represent fundamental challenges in image processing with many real-world applications. To address practical demands, current solutions can be categorized into two main strategies: 1) sequential method; and 2) joint method. However, sequential methods have the disadvantage of error accumulation as there is information loss between multiple individual models. Recently, the academic community began to make some attempts to tackle this problem through end-to-end joint methods. Most of them ignore that different regions of noisy images have different characteristics. To solve these problems, in this paper, our proposed signal-to-noise ratio~(SNR) aware joint solution exploits local and non-local features for image compression and denoising simultaneously. We design an end-to-end trainable network, which includes the main encoder branch, the guidance branch, and the signal-to-noise ratio~(SNR) aware branch. We conducted extensive experiments on both synthetic and real-world datasets, demonstrating that our joint solution outperforms existing state-of-the-art methods. 
\end{abstract}
\section{Introduction}
Image denoising, a vital task in low-level computer vision, plays a crucial role in numerous high-level applications. The recent years have marked remarkable advancements in image denoising, primarily attributed to the adoption of sophisticated deep neural networks.
In real-world scenarios, lossy image compression is essential for efficient media storage and transmission.
In recent years, learning-based lossy image compression methods~\cite{he2021checkerboard}\cite{qian2022entroformer}\cite{zhu2022unified} have made significant progress, outperforming traditional standards on performance metrics such as peak signal-to-noise ratio (PSNR) and multi-scale structural similarity index (MS-SSIM).
Given that the majority of current compression approaches are tailored for general images. These compressors perceive noise as essential information and allocate bits explicitly to preserve it, disregarding the fact that noise is typically undesired by ordinary users, abbreviated as ``bits misallocation problem''.
As mentioned in previous research~\cite{al1998lossy}\cite{ponomarenko2010lossy}, image noise can degrade image compression quality.
\par
In various real-world systems such as autonomous driving and visual surveillance, there is a requirement for lossy noisy image compression, while minimal research has been conducted in the academic community regarding this practical subject.
Existing engineering solutions can be classified into two manners: ``Compress before Denoise'' and ``Denoise before Compress''.
However, those sequential solutions have the disadvantage of error accumulation as there is information loss between multiple individual models.
More specifically, the ``Compress before Denoise'' solution tends to suffer from bits misallocation. In the case of the ``Denoise before Compress'' solution, the rate-distortion performance of the compression method is predominantly constrained by the denoising approach. The quality of the denoised image will be further degraded after using the lossy codec.
\par
In addition to the aforementioned sequential solutions, the academic community has recently made some attempts to tackle this problem through end-to-end joint manners~\cite{testolina2021towards}\cite{cheng2022optimizing}\cite{huang2023narv}. Most of them ignore that the different regions in the noisy image have different characteristics.
In this paper, we present a key insight that different regions within a noisy image may exhibit distinct characteristics.
More specifically, regions with complex textures are significantly impacted by noise, whereas areas with fewer textures in the same image may experience only mild interference.
To achieve superior reconstruction results, it is crucial to adaptively consider the distinct characteristics of various regions within the noisy images during the encoding and denoising procedure.
\par
To achieve spatially varying image denoising and compression simultaneously, we investigate the relationship between signal and noise in image space through the exploration of signal-to-noise ratio~(SNR)~\cite{chandler2007vsnr}. 
Regions with lower SNR often exhibit unclear details and substantial noise interference. To address this, we leverage non-local image information across a long spatial range for effective image denoising and compression. Conversely, regions with relatively higher SNR experience fewer noise interferences, making local image information generally adequate.
Based on these considerations, in this work, our proposed joint solution is to collectively exploit local and non-local information for image compression and denoising. 
We formulate an end-to-end trainable architecture with three branches: 1) the main encoder branch extracts compressed domain features, 2) the teacher guidance branch generates two-level guiding features, 3) the signal-to-noise ratio~(SNR) aware branch captures local and non-local features.
The local/non-local features are fused with the compressed domain features, producing denoised features that enable simultaneous image compression and denoising. 
Finally, denoised images are reconstructed through the main decoder. 
In summary, the contributions of this work are as follows:
\begin{itemize}
\item We propose a new signal-to-noise~(SNR) aware framework that extracts local and non-local features for joint image compression and denoising with the SNR prior.
\item The end-to-end trainable three-branch architecture empowers the joint solution to obtain high-quality reconstructed images at low bits-per-pixel~(BPP).
\item We conduct extensive experiments on both synthetic and real-world datasets, demonstrating that our proposed SNR-aware joint framework consistently outperforms state-of-the-art methods.
\end{itemize}

\section{Related Works}
\subsection{Learning-based Lossy Image Compression}
The works~\cite{balle2017end}\cite{balle2018variational}\cite{theis2017lossy} initially used neural networks for end-to-end image compression and inspired many subsequent learning-based image compression methods.
Subsequent researchers have conducted extensive research on the three components typically involved in lossy image compression: transformation, quantization, and entropy coding.
Some previous research focuses on quantization, \textit{e.g.}, Dumas~\textit{et al.}~\cite{dumas2018autoencoder} aimed to learn different quantization step sizes for various latent representations.
Some works focus on the transform, \textit{e.g.}, residual block~\cite{theis2017lossy}, attention module~\cite{cheng2020learned}, and transformer-based architecture~\cite{qian2022entroformer}.
Some works aim to improve the efficiency of entropy coding, \textit{e.g.}, channel-wise entropy model~\cite{minnen2020channel}, checkerboard context model~\cite{he2021checkerboard}, and multivariate Gaussian mixture model~\cite{zhu2022unified}.
However, the mutual influence with image processing tasks is not considered in the design of most existing learning-based compression methods. Combined with image processing tasks in the practical systems, this can lead to unsatisfactory image quality and suboptimal subsequent visual tasks.
\subsection{Image Denoising}
Image denoising is an image restoration task that has been studied for a long time, and many traditional methods have been proposed in the past decades. 
Those traditional methods~\cite{chambolle2004algorithm}\cite{buades2005non}\cite{gu2014weighted} are typically designed by exploiting certain signal structures or noise characteristics that do not rely on data-driven learning.
As neural networks are rapidly developing, more and more methods are utilizing neural networks to improve image-denoising performance. 
There have been some works on adapting synthetic dataset approaches to real-world scenarios~\cite{guo2019toward}\cite{kim2020transfer}.  
Recently, the performance has been further improved by some state-of-the-art methods~\cite{jiang2023dynamic}\cite{chen2023masked}. 
However, many learning-based methods are designed without considering the effects on other tasks~(\textit{e.g.}, image compression). This is suboptimal in the real-world image processing pipeline which typically involves multiple tasks, such as image denoising and image compression.
\subsection{Joint Solutions}
Some joint solutions~\cite{ehret2019joint}\cite{xing2021end} have been verified as an effective alternative to sequential ones with promising results in the image processing pipeline. 
Testolina \textit{et al.}~\cite{testolina2021towards} introduced the idea of denoising into image decompression by denoising the latent representations during the decoding procedure. 
Cheng \textit{et al.}~\cite{cheng2022optimizing} designed a plug-in module that can perform denoising during image encoding processing.
Alvar \textit{et al.}~\cite{ranjbar2022joint} used scalable coding to divide the bitstream into useful image information and noise.
Huang \textit{et al.}~\cite{huang2023narv} proposed the Noise-Adaptive ResNet VAE~(NARV) to handle both clean and noisy images by a single image compression model.
However, the above-mentioned methods of joint image compression and image denoising ignored that the different regions in the noisy image have different characteristics. These characteristics can be utilized to efficiently identify useful content and noise in the image, allowing for higher image compression and image denoising performance.

\begin{figure*}[t]
\centering
\includegraphics[width=0.86\linewidth, height=5cm]{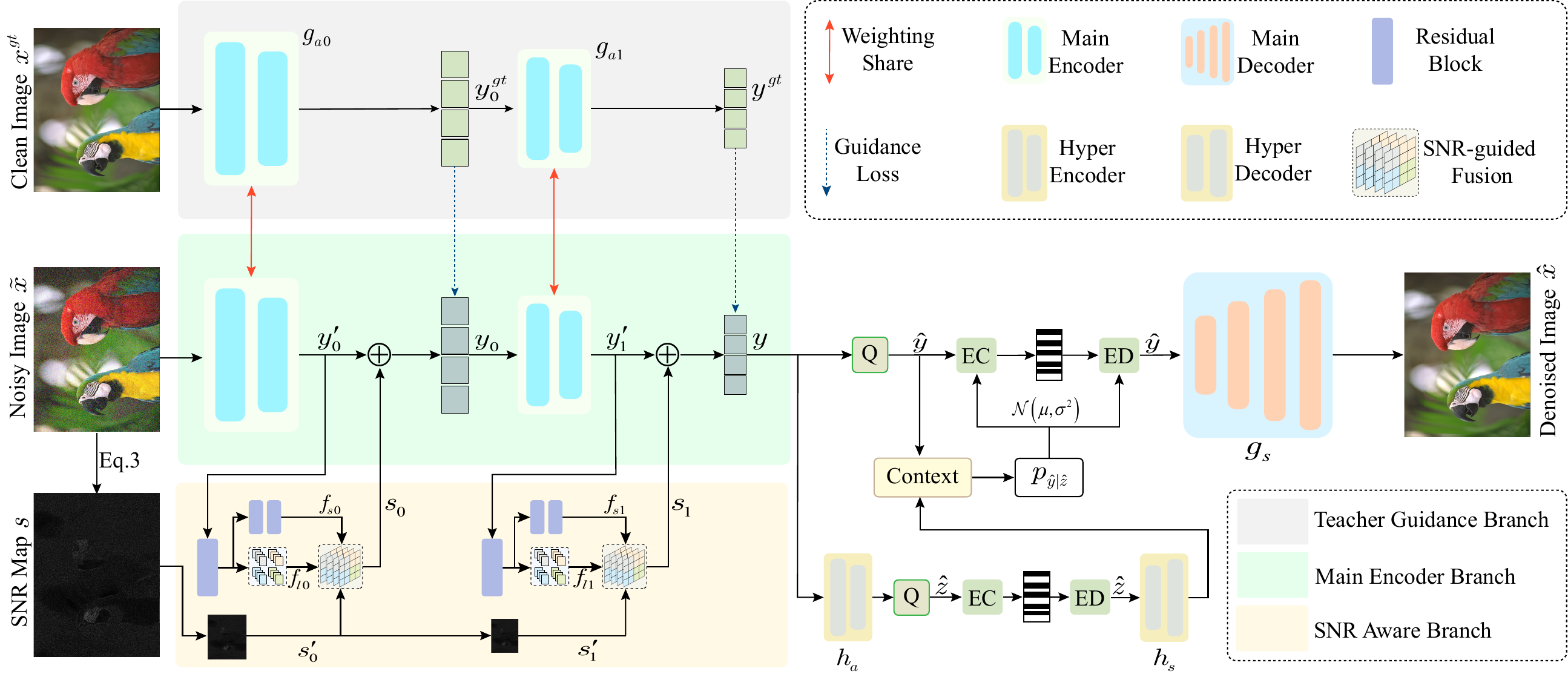}
\caption{The network architecture of proposed SNR-aware joint solution of image compression and denoising. The architecture contains three branches, ``Teacher Guidance Branch", ``Main Encoder Branch" and ``SNR Aware Branch", in the left half of the figure. The right half of the figure contains the main decoder, entropy models, context model, and hyper encoder/decoder commonly used in recent learning-based compression method~\cite{cheng2020learned}. Note that the ``Teacher Guidance Branch" is for training only, the ``Main Encoder Branch" and ``SNR Aware Branch" are activated during training of the entire network and used for inference. $\bigoplus$ denotes the addition by element.}
\label{Joint_share_SNR_aware_architecture} 
\end{figure*}
\section{Proposed Method}
\subsection{Framework}
\textbf{Overall workflow.} Fig.~\ref{Joint_share_SNR_aware_architecture} depicts the overview of the network architecture of our proposed signal-to-noise ratio~(SNR) aware joint image compression and denoising framework. 
The joint framework contains a novel three-branch design for the training procedure. 
The noisy image $\tilde{x}$ is transformed to the denoised compressed domain latent representation $y$ through the main encoder $g_{a0}$ and $g_{a1}$ with the signal-to-noise ratio~(SNR) aware branch. 
Then, quantization of $y$ to the discrete denoised compressed domain latent representation $\hat{y}$ is performed by the quantizer Q. 
During the training procedure, we replace the non-differentiable quantization operation with the addition of uniform noise $U(- \frac{1}{2},  \frac{1}{2})$ to the denoised compressed domain latent representation $y$. 
During the testing procedure, we round the denoised compressed domain latent representation $y$, directly. 
The hyper-prior scale entropy model~\cite{cheng2020learned}\cite{minnen2018joint} is utilized in our proposed framework to effectively estimate the distribution $p_{\hat{y}|\hat{z}} \sim \mathcal{N}(\mu, \sigma^{2})$ of the discrete denoised compressed latent representation $\hat{y}$. 
The Gaussian parameters $\mu$ and $\sigma$ are generated by the entropy model to support the entropy encoding/decoding procedure. 
The range asymmetric numeral system~\cite{duda2013asymmetric} is used to losslessly encode discrete denoised features $\hat{y}$ and latent representation $\hat{z}$ into bitstreams.
The latent representation $z$ is quantized using the same strategy as the denoised compressed latent representation $y$. The factorized entropy model~\cite{balle2017end} is used to estimate the distribution $p_{\hat{z}}$ of the discrete latent representation $\hat{z}$. The entropy decoding yields the discrete denoised compressed latent representation $\hat{y}$, which is then fed into the main decoder $g_{s}$ to reconstruct the denoised image $\hat{x}$.
\\
\textbf{Three branch architecture.} 
As shown in Fig.~\ref{Joint_share_SNR_aware_architecture}, our proposed joint solution consists of three branches: 1) the main encoder branch, 2) the teacher guidance branch, and 3) the signal-to-noise ratio~(SNR) aware branch.
During the training procedure, the noisy image $\tilde{x}$ and the corresponding clean image $x^{gt}$ are fed into three branches, respectively.
The main encoder branch combines the compressed domain features~($y_{0}^{\prime}/y_{1}^{\prime}$) with the local/non-local information~($s_{0}$  and $s_{1}$) generated by the SNR-aware branch to obtain the denoised compressed domain latent representation~($y_{0}/y_{1}$); the calculation process is abbreviated as:~$\textit{SNR}(\cdot, \cdot)$.
The SNR aware branch generates local/non-local features~($s_{0}/s_{1}$) through SNR map $s$. The SNR map $s$ is achieved through a simple yet effective non-learning-based denoising operation~(refer Eq.\ref{SNR_map_calculate}). 
In addition, the clean image $x^{gt}$ is fed to the 
teacher guidance branch~(training procedure only) to generate two-level guiding features~($y_{0}^{gt}$/$y_{1}^{gt}$) by the main encoder blocks~($g_{a0}$/$g_{a1}$), respectively.
The two-level guidance and denoised compressed latent representation $y_{0}^{gt}$/$y^{gt}$ and $y_{0}$/$y$ can be expressed by formulas:
\begin{equation}
\begin{aligned}
&y_{0}^{gt}=g_{a0}(x^{gt}), \quad y_{0}=g_{a0}(\tilde{x})+\textit{SNR}(s_{0}, g_{a0}(\tilde{x})), \\
&y^{gt}=g_{a1}(y_{0}^{gt}), \quad y=g_{a1}(y_{0})+\textit{SNR}(s_{1}, g_{a1}(y_{0})).
\end{aligned}
\end{equation}
During the training procedure, the $L_{1}$ distance between denoised and guidance latent representation is minimized:
\begin{equation}
\mathfrak{L}_{g} = ||y_{0}^{gt}-y_{0}||_{1} + ||y^{gt}-y||_{1}.
\label{guidance_loss}
\end{equation}

\subsection{SNR-Aware Denoising in Compressed Features}
The signal-to-noise ratio~(SNR) map $s \in \mathbb{R}^{h \times w}$ is obtained by converting the noisy image $\tilde{x}  \in \mathbb{R}^{3 \times h \times w}$ to the grayscale image $\dot{x}  \in \mathbb{R}^{h \times w}$ and then using the following formulas:
\begin{equation}
	\ddot{x} = kernel(\dot{x}), \quad n=abs(\dot{x} - \ddot{x}), \quad s = \frac{\ddot{x}}{n},
	\label{SNR_map_calculate}
\end{equation}
where $kernel(\cdot)$ represents the operation of averaging local pixel groups, while $abs(\cdot)$ indicates the application of the absolute value function. 
As shown in Fig.~\ref{Joint_share_SNR_aware_architecture}, we utilize the ``Residual Block'' to extract features. These features are subsequently processed separately by the long- and short-range modules, resulting in the generation of non-local features ($f_{l0}/f_{l1}$) and local features ($f_{s0}/f_{s1}$).
The two-level local and non-local features are fused. It is illustrated in ``SNR-guided Fusion'' of Fig.~\ref{Joint_share_SNR_aware_architecture} and is calculated as follows:
\begin{equation}
 \begin{split}
 	&s_{0}=f_{s0} \times s_{0}^{\prime} + f_{l0} \times (1 - s_{0}^{\prime}),\\
 	&s_{1}=f_{s1} \times s_{1}^{\prime} + f_{l1} \times (1 - s_{1}^{\prime}),
 \end{split}
\end{equation}
where $s_{0}^{\prime}$ and $s_{1}^{\prime}$ are resized from SNR map $s$ according to the shape of local/non-local features~($f_{s0}$/$f_{s1}$/$ f_{l0}$/$f_{l1}$).
The SNR-aware fusion features, represented as $s_{0}$/$s_{1}$, are added to the main features~($y_{0}^{\prime}$/$y_{1}^{\prime}$) to generate the denoised compressed latent representation~($y_{0}$/$y$).
\par
Short-range modules are constructed using a few residual blocks.
Long-range modules are constructed by SNR-aware transformers with self-attention blocks. In the long-range modules, the feature maps $F \in \mathbb{R}^{h\times w \times C}$ are divided into $t$ feature patches, represented as $F^{i} \in \mathbb{R}^{m\times m \times C}$, $i=\{1,...,t\}$. The relationship between patch size $m\times m$ and the entire feature map can be described as $t=\frac{h}{m}\times \frac{w}{m}$.
The feature patches $F^{1},..., F^{t}$ is flatten into 1D vectors, and processed by: 
\begin{equation}
\begin{split}
&f^{0}=[F^{1}, F^{2}, ..., F^{t}],\\
&q^{i} = k^{i} = v^{i} = LN(f^{i-1}),\\
&\hat{f^{i}}=MSA(q^{i}, k^{i}, v^{i})+f^{i-1},\\
&f^{i}=FFN(LN(\hat{f^{i}}))+\hat{f^{i}},\\
&[\mathfrak{F}^{1}, \mathfrak{F}^{2}, ..., \mathfrak{F}^{t}]=f^{r}, i\in\{1,2, ..., r\}.
\end{split}
\end{equation}
Where $LN(\cdot)$ is layer normalization. $f^{i}$ is the result calculated by $i\text{-th}$ transformer block. $MSA(\cdot)$ is the muti-head self-attention module~\cite{vaswani2017attention}. $FFN(\cdot)$ is the feed-forward network~\cite{vaswani2017attention}. $q^{i}$, $k^{i}$, and $v^{i}$ represent query, key, and value vectors, respectively. $r$ is the number of layers in the transformer. $\mathfrak{F}^{1}, \mathfrak{F}^{2}, ..., \mathfrak{F}^{t}$ are the outputs of the transformer block which can be reshaped into 2D feature map~($f_{l0}/f_{f_{l1}}$).

\subsection{Rate-Distortion Optimization}
In terms of image compression, we aim to minimize the length of the stored bitstreams; while for image denoising, we aim to minimize the difference between the decoded image $\hat{x}$ and the paired clean image $x^{gt}$. Thus in this joint task, we can use the rate-distortion~(RD) objective function as follows:
\begin{equation}
 \mathfrak{L}_{rd}=\mathfrak{R}(\hat{y}) + \mathfrak{R}(\hat{z})+\lambda_{d} \cdot \mathfrak{D}(\hat{x}, x^{gt}),
\end{equation}
where the $\mathfrak{D}(\hat{x}, x^{gt})$ is the distortion between decoded denoising image $\hat{x}$ and clean image $x^{gt}$.  Similar to previous works~\cite{balle2018variational}\cite{cheng2020learned}, the $\mathfrak{R}(\cdot)$ represents the compression level of the discrete latent representations $\hat{y}$, $\hat{z}$~(as shown in Fig.~\ref{Joint_share_SNR_aware_architecture}) , defined as follows:
\begin{small}
\begin{equation}
\mathfrak{R}(\hat{y})=\mathbb{E}_{q_{\hat{y}}}\bigl[\text{-}\log p_{\hat{y}|\hat{z}}(\hat{y}|\hat{z})\bigr], \mathfrak{R}(\hat{z})=\mathbb{E}_{q_{\hat{z}}}\bigl[\text{-}\log p_{\hat{z}|\theta}(\hat{z}|\theta)\bigr].
\end{equation}
\end{small}
$\lambda_{d}$ denotes the weighting coefficient, which is the trade-off between compression levels and distortion.
Considering guidance loss Eq.\ref{guidance_loss}, the objective function of the entire network is:
\begin{equation}
\begin{aligned}
\mathfrak{L}= \mathfrak{L}_{rd}+\lambda_{g} \cdot \mathfrak{L}_{g},
\end{aligned}
\end{equation}
where the parameter $\lambda_{g}=3$ represents the weight factor assigned to the guidance loss. 

\section{Experiments}
\subsection{Datasets and Implementation details}
\textbf{Synthetic datasets.} The training and validation dataset is Flicker 2W~\cite{liu2020unified}. Images smaller than 256 pixels are excluded, and approximately 200 images are chosen for validation. During the training procedure, randomly cropped patches with a resolution of $256 \times 256$ pixels are used to optimize the joint framework. The CLIC~\cite{CLIC2020} Professional Validation and Kodak~\cite{kodak} datasets are employed for testing. For more details about noise synthesis, please refer to the supplementary material.\\
\textbf{Real-world datasets.} The proposed joint framework is trained using the SIDD Medium~\cite{abdelhamed2018high} dataset, consisting of 320 pairs of noisy-clear sRGB images in the training set. Subsequently, real-world noisy images are utilized in the testing procedure. The models undergo validation using the 1280 patches from the SIDD validation set and are tested on the SIDD benchmark patches, with the results submitted for testing on the SIDD website.
\\
\textbf{Implementation details.} Our implementation relies on PyTorch and the open-source CompressAI library. The networks are optimized using the Adam optimizer with a mini-batch size of 16, trained for approximately 600 epochs on RTX 3090 GPUs. The initial learning rate is set as $5 \times 10^{-5}$ and decayed by a factor of 0.1 at epochs 450 and 500. We set a loss cap for each model, ensuring that the network skips optimizing a mini-step if the training loss exceeds the specified threshold. We train MSE models across 6 qualities, with $\lambda_{d}$ selected from the set \{0.0018, 0.0035, 0.0067, 0.0130, 0.0250, 0.0483\}; the corresponding $\lambda_{d}$ values for MS-SSIM models are chosen from \{4.58, 8.73, 31.73, 60.50\}. For improved visualization, the MS-SSIM is converted to decibels using the formula $-10 \log_{10}(1-\text{MS-SSIM})$.
\begin{figure}[t]
	\centering
	\includegraphics[width=0.49\linewidth]{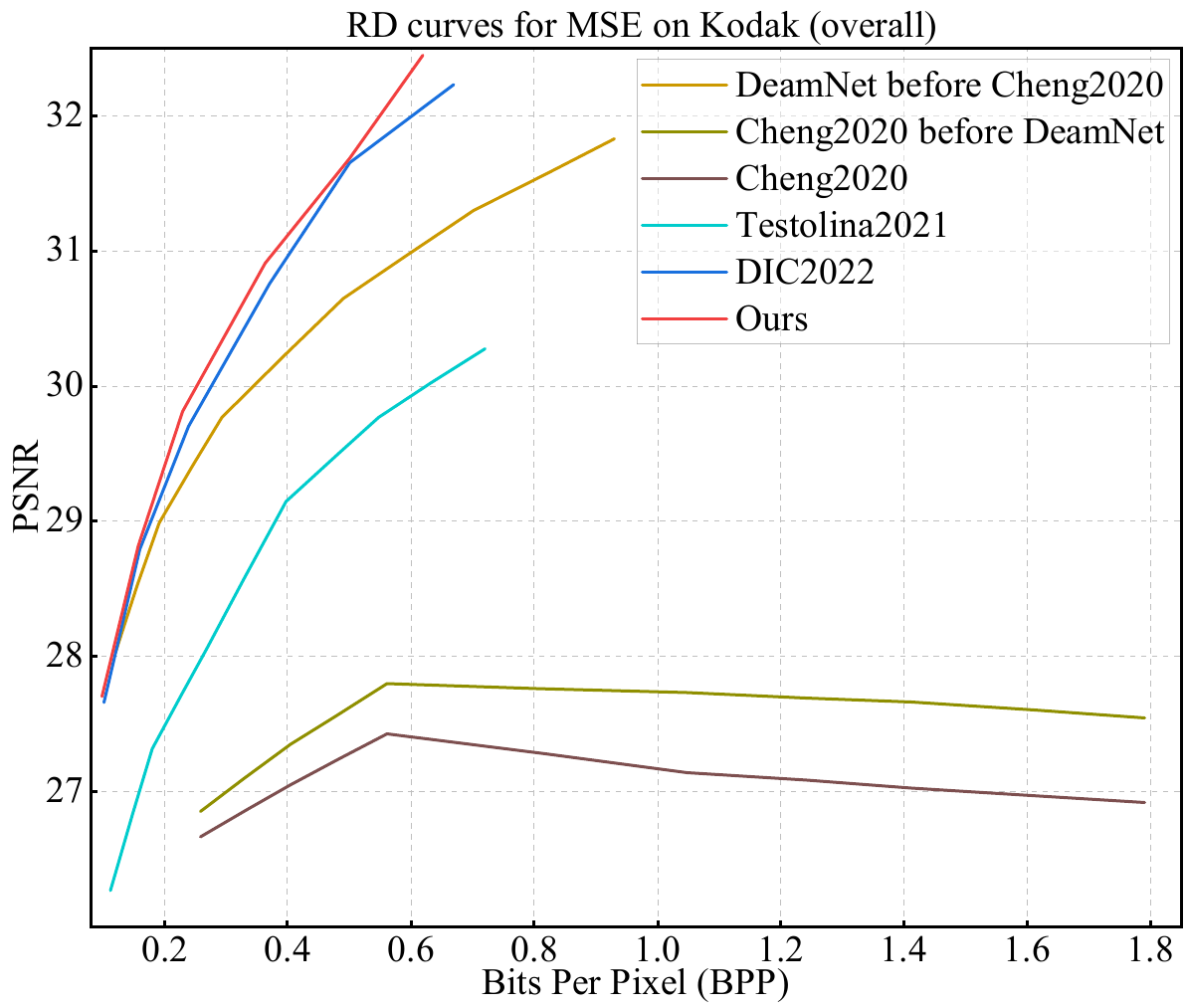}\hfill
	\includegraphics[width=0.49\linewidth]{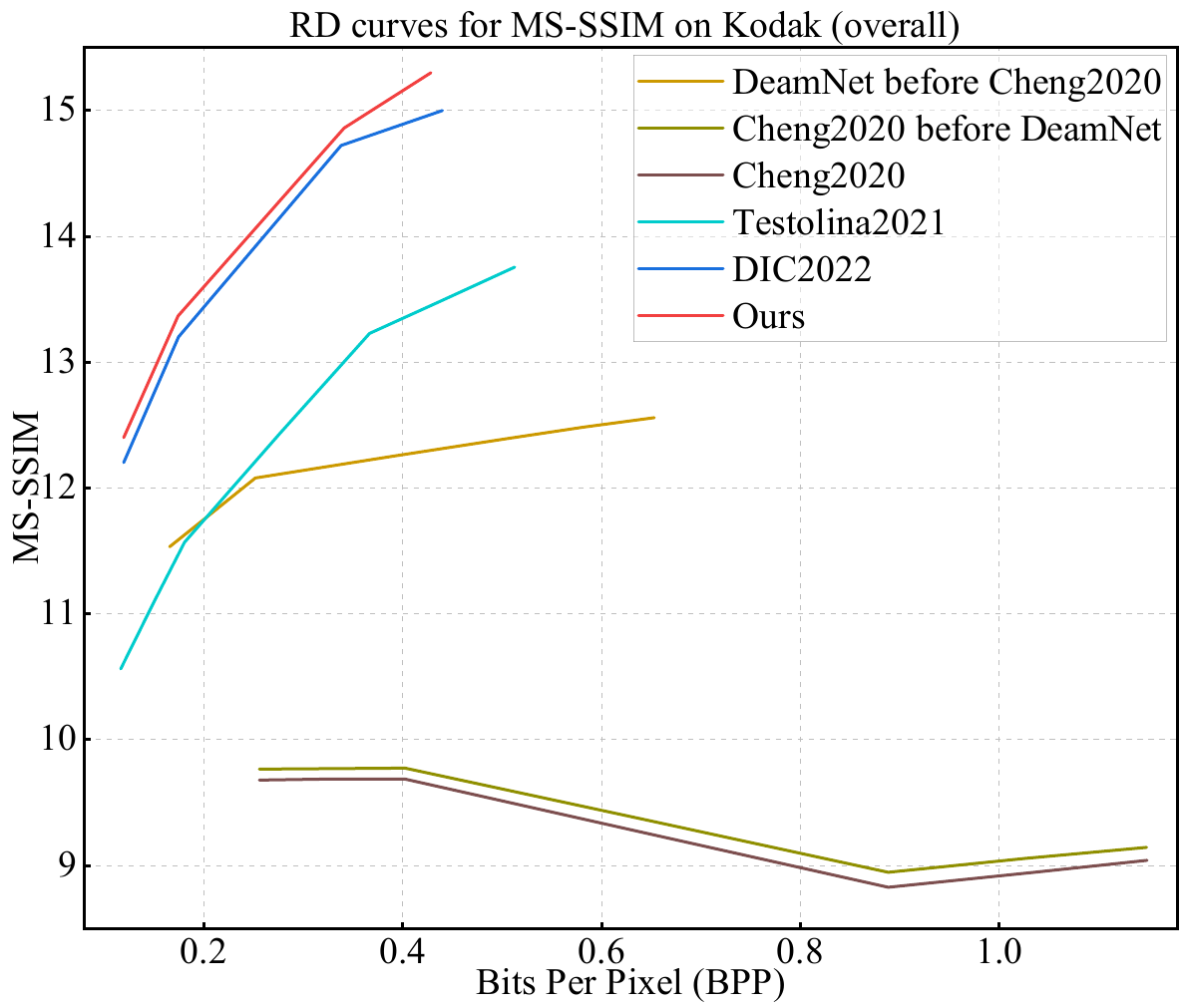}\\
	\includegraphics[width=0.49\linewidth]{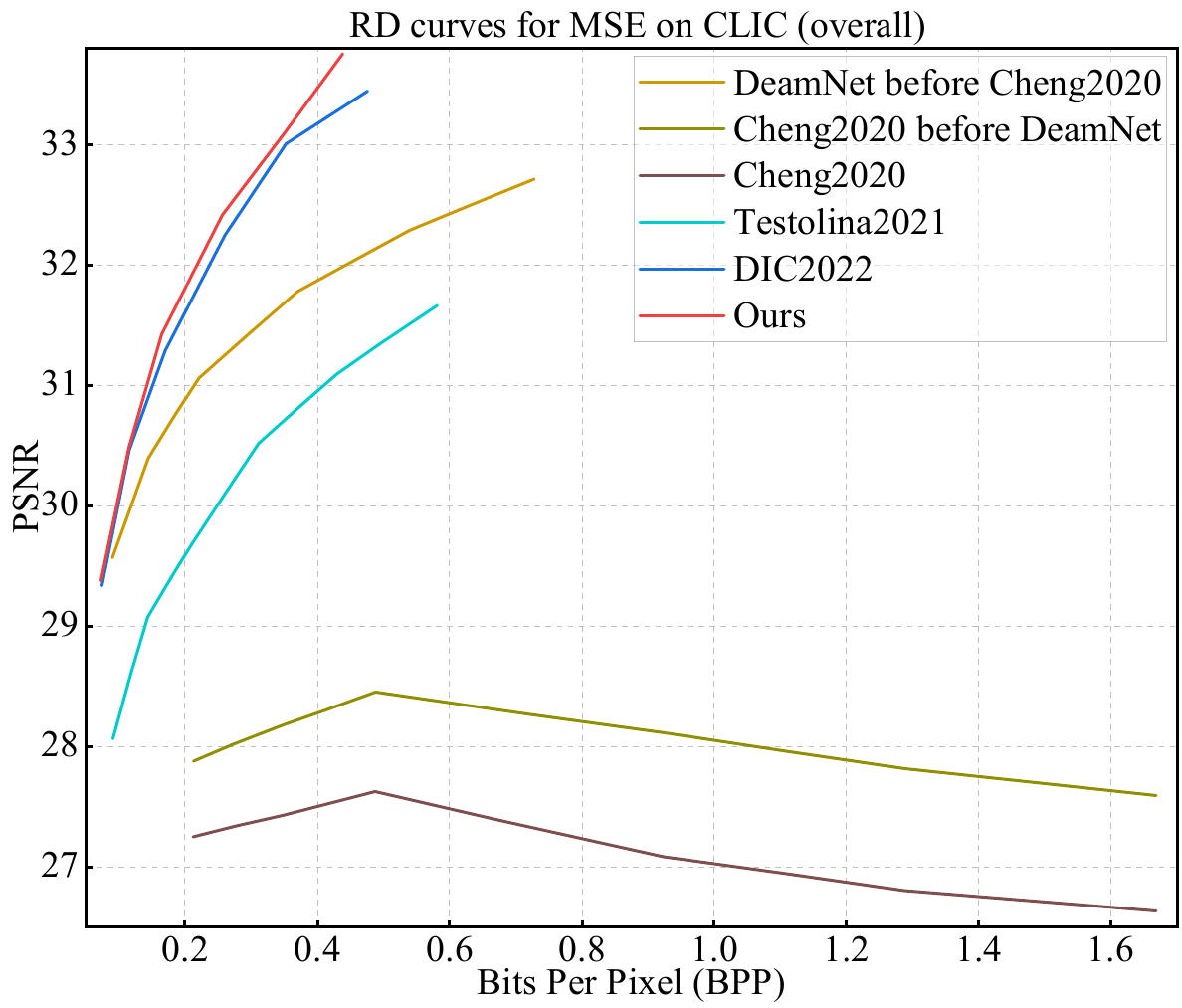}\hfill
	\includegraphics[width=0.49\linewidth]{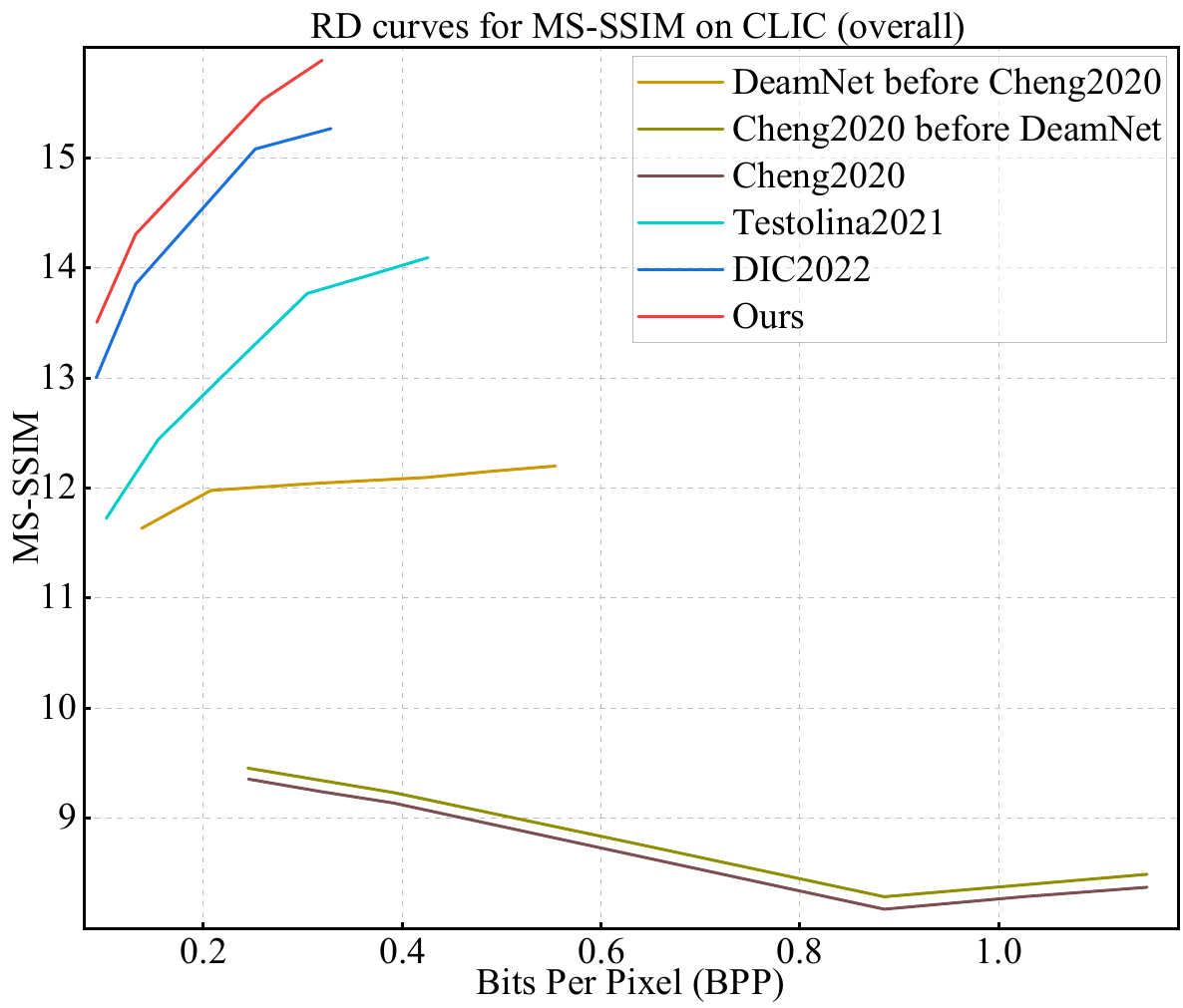}\vspace{-0.2cm}
	\caption{Overall RD curves for the Kodak and CLIC datasets across all noise levels. Our proposed joint solution, indicated by red curves, exhibits superior RD performance compared to pure compression, sequential, and joint methods.}\vspace{-0.3cm}
	\label{Kodak_overall}
\end{figure}

\begin{figure}[t]
	\centering
	\includegraphics[width=0.49\linewidth]{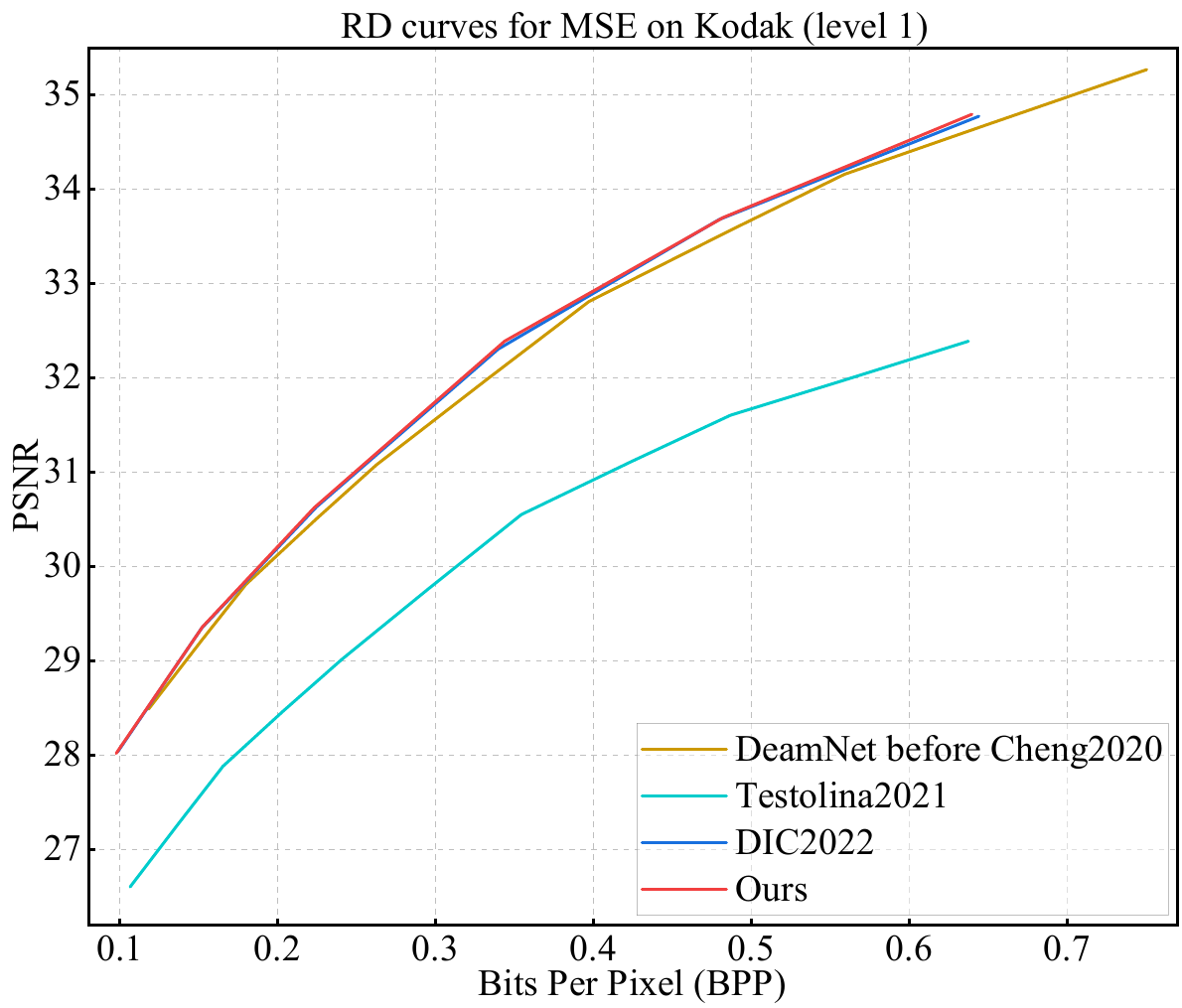}\hfill
	\includegraphics[width=0.49\linewidth]{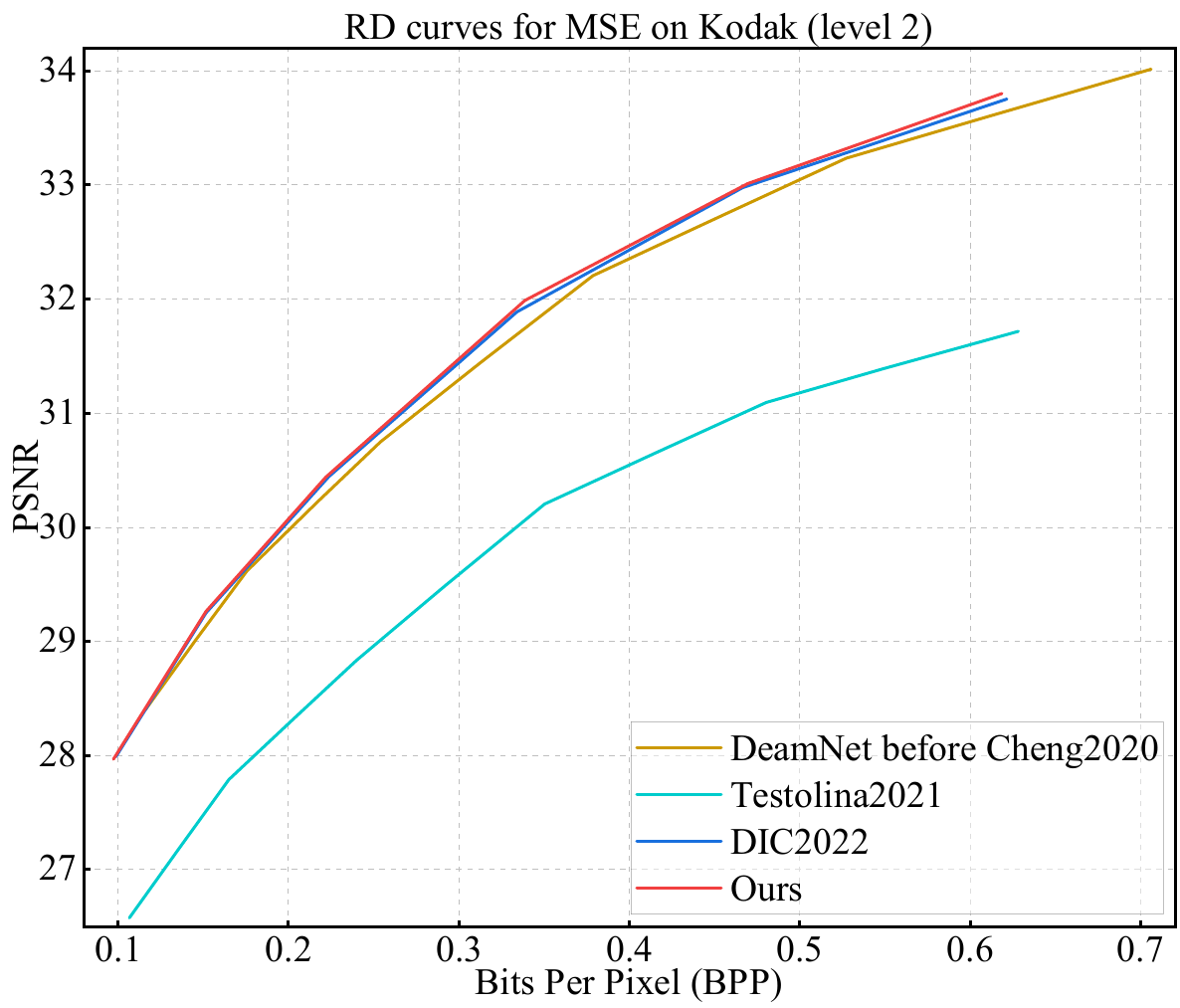}\\
	\includegraphics[width=0.49\linewidth]{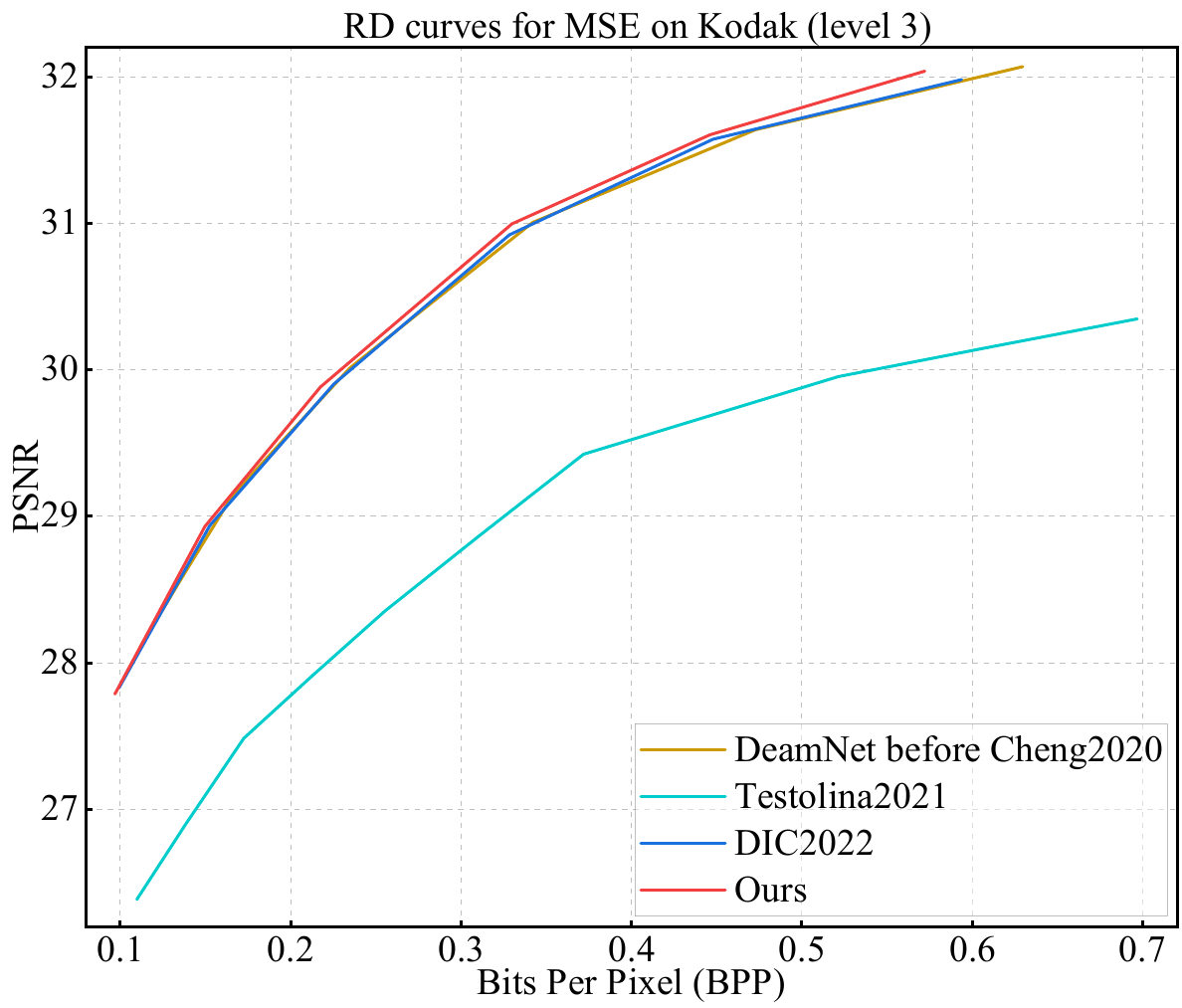}\hfill
	\includegraphics[width=0.49\linewidth]{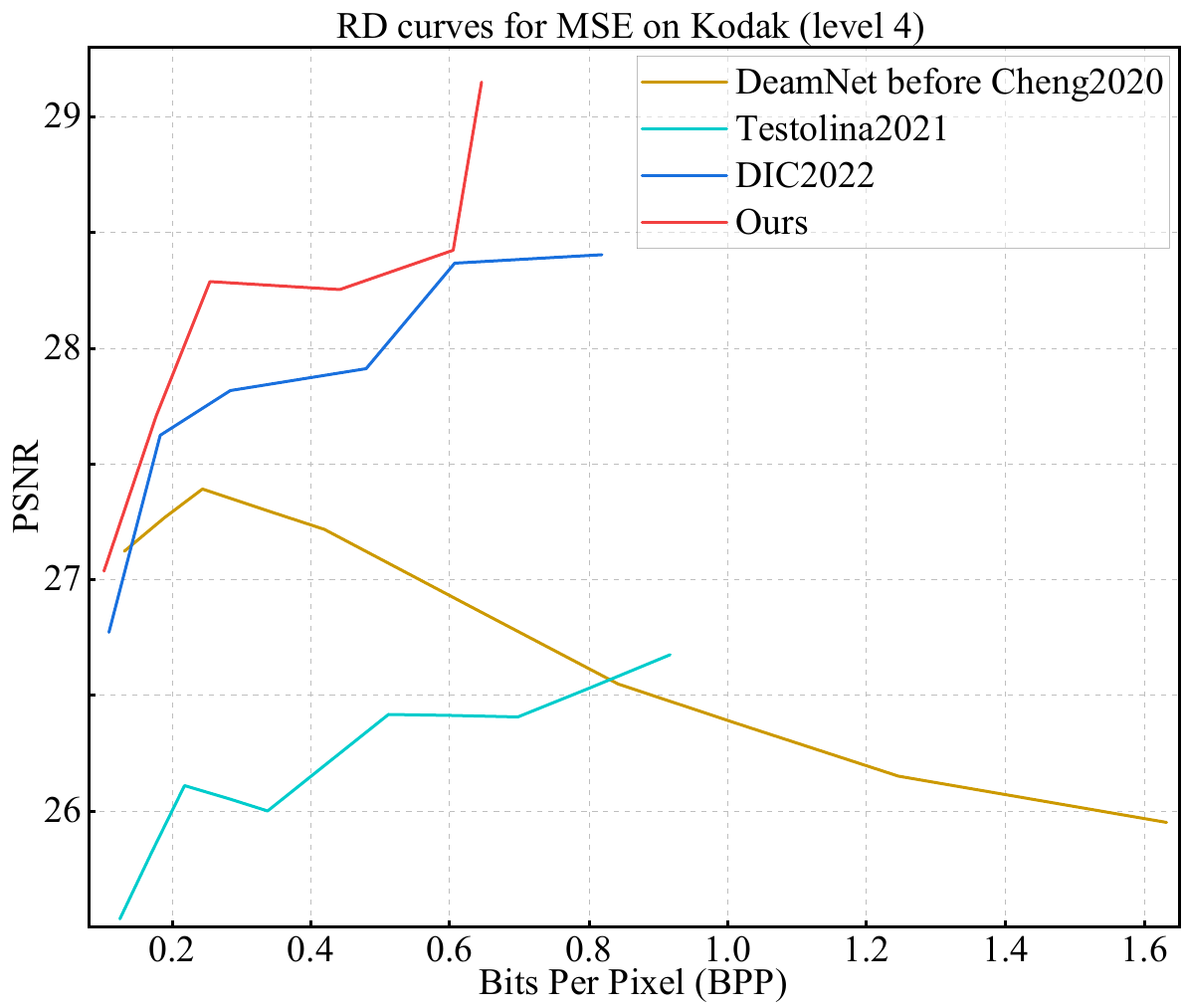}\vspace{-0.2cm}
	\caption{RD curves on the Kodak dataset at various noise levels. Our method surpasses both sequential and joint methods, particularly at the high noise level. This denotes that the proposed SNR-aware branch efficiently captures valuable information through a combination of local and non-local features.}
	\label{kodak_individual}
\end{figure}
\subsection{Rate-Distortion Performance}
\begin{figure}[h]
\centering
\includegraphics[width=0.49\linewidth]{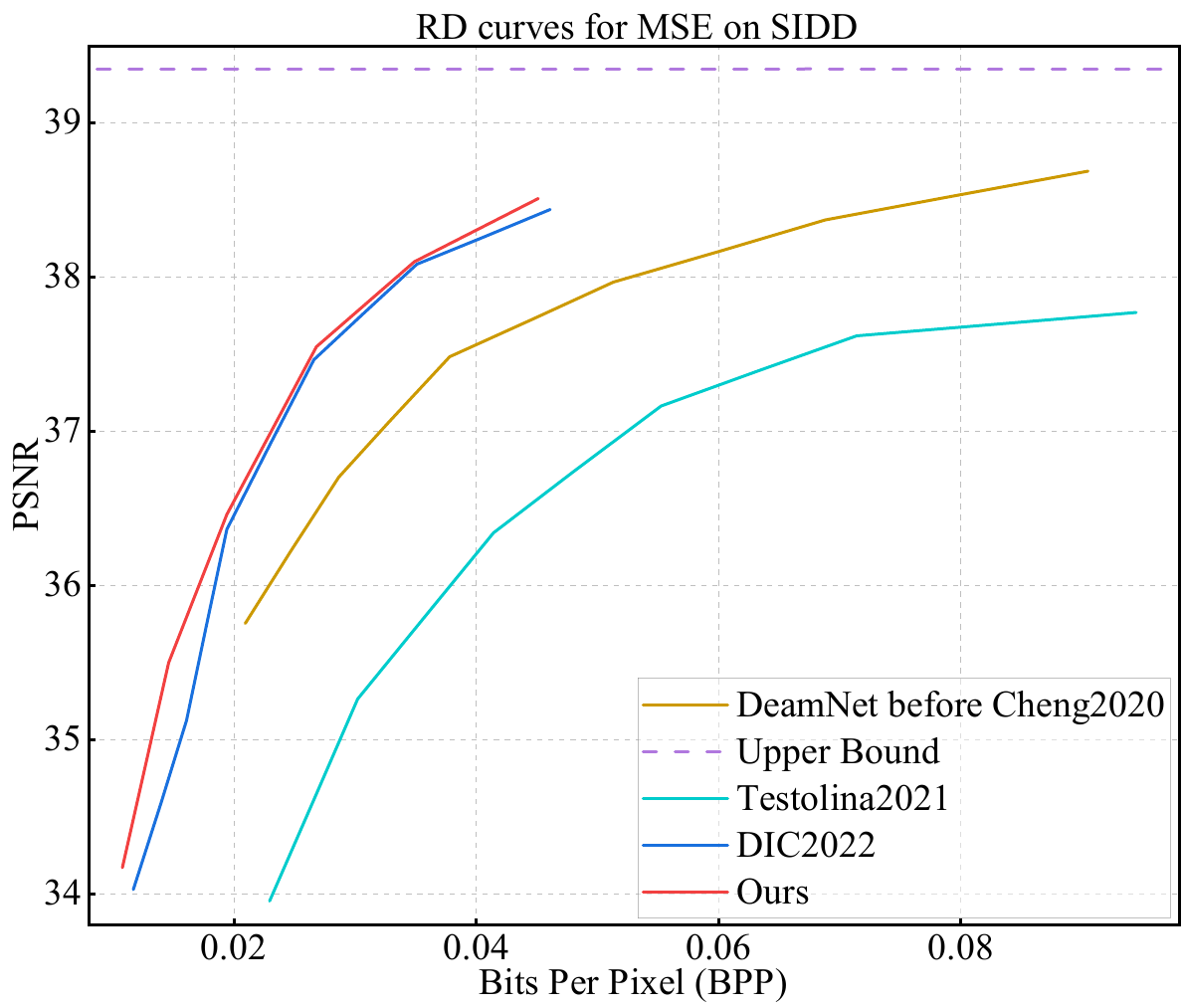}\vspace{-0.2cm}
\caption{RD performance curves optimized by MSE aggregated on SIDD. Our proposed method achieved the best RD performance. This indicates that our method is robust on real noisy images. The purple dotted line serves as a reference for the DeamNet~\cite{ren2021adaptive} ideal case without compression.}\vspace{-0.4cm}
\label{SIDD_PSNR} 
\end{figure}

Our proposed joint solution compares with the five methods: 
1) ``DIC2022'': the state-of-the-art joint method~\cite{cheng2022optimizing}; 2) ``Testolina2021'': an early joint method~\cite{testolina2021towards}; 
3) ``Cheng2020 before DeamNet'': sequential method of Cheng2020~\cite{cheng2020learned} and DeamNet~\cite{ren2021adaptive}; 4) ``DeamNet before Cheng2020'': sequential method of DeamNet~\cite{ren2021adaptive} and Cheng2020~\cite{cheng2020learned};
5) ``Cheng2020'': the compression method~\cite{cheng2020learned} is used on noise-to-clean image pairs directly. RD results are obtained from the CompressAI evaluation platform, the official SIDD website, or provided in the corresponding paper. 
More experimental results, both quantitative and visual, are provided in the supplementary material.\\
\textbf{Synthetic noise~(overall).} In Fig.~\ref{Kodak_overall}, we present overall RD curves for both MSE and MS-SSIM methods, covering all four noise levels, as evaluated on the Kodak and CLIC datasets, respectively. Our proposed joint solution, depicted by the red RD curves, outperforms the pure compression, sequential, and joint methods in terms of overall performance.
\\
\textbf{Synthetic noise~(individual).} To delve deeper into the impact of various noise levels, Fig.~\ref{kodak_individual} illustrates RD curves for MSE models on the Kodak dataset at individual noise levels. Our proposed joint method is better than the sequential and joint methods, especially at the high noise level. 
The SNR in an image is usually low when the noise level is high. Therefore, it is difficult to achieve high-quality reconstructed images using only local features. Thanks to our proposed SNR-aware branch through local and non-local feature fusion, we can effectively be aware of different SNR regions in the image and then obtain a higher-quality reconstructed image. More experimental results are provided in the supplementary material.
\\
\textbf{Real-world noise.} In Fig.~\ref{SIDD_PSNR}, we present the RD curves optimized for MSE on the SIDD test set with real-world noise. The purple dashed line indicates the results of the pure denoising model DeamNet~\cite{ren2021adaptive} to show the upper bound of the denoising performance without compression~($\text{BPP}=24$). The results demonstrate the effectiveness of our proposed joint framework, proving its efficiency not only on synthetic datasets but also on images containing real-world noise.

\vspace{-0.3cm}
\section{Conclusion}
\vspace{-0.3cm}
We introduce an innovative signal-to-noise ratio~(SNR) aware joint solution to improve the capability of lossy image compression for noisy images. 
Thanks to the end-to-end trainable three-branch architecture, the joint solution can achieve high-quality denoised images at low BPP.
The different regions in the noisy image have different characteristics, these characteristics can be utilized in the local and non-local information fusion procedure efficiently.
Local and non-local information obtained by the SNR-aware branch would be fused with the compressed features to generate compressed denoised features. 
Finally, the denoised image can be obtained by decoding the compressed denoised features directly. 
The experimental results show that our proposed SNR-aware joint solution surpasses sequential and joint methods on both synthetic and real-world datasets.

\vspace{-0.4cm}
\section{Acknowledgments}
This work was supported by the National Natural Science Foundation of China under Grant 62301228, 62176100.
The computation is completed in the HPC Platform of Huazhong University of Science and Technology.

\bibliographystyle{IEEEbib}
\bibliography{reference}

\appendix
%
\clearpage
\newpage
\begin{center}
\section*{Summary}
\end{center}
This supplementary material is organized as follows:
\begin{itemize}
    \item Section~\ref{sec1} describes more details about noise synthesis.
    \item Section~\ref{sec2} provides more RD performance comparison results on the CLIC Professional Validation~\cite{CLIC2020} dataset.
    \item Section~\ref{sec3} shows our proposed method is used for general image compression. It indicates our framework is also effective for compressing generic images.
    \item Section~\ref{sec4} provides visualization results.
\end{itemize}

\section{Noise Synthesis}
\label{sec1}
We use the same noise synthesis strategy as prior research~\cite{mildenhall2018burst}. 
In the training procedure, we acquire the noise parameter $\delta_{r}$ and shot noise parameter $\delta_{s}$ through uniform sampling from the intervals $[10^{-3}, 10^{-1.5}]$ and $[10^{-4}, 10^{-2}]$, respectively. 
During testing and validation procedures, we utilize the four pre-determined parameter pairs $\{\text{Gain}\propto 1=(10^{-2.1},10^{-2.6}), \text{Gain}\propto 2=(10^{-1.8},10^{-2.3}), \text{Gain}\propto 4=(10^{-1.4},10^{-1.9}), \text{Gain}\propto 8=(10^{-1.1},10^{-1.5})\}$.
It is worth noting that, the network is unaware of the $\text{Gain}\propto 4=(10^{-1.4},10^{-1.9})$~(slightly noisier) and $\text{Gain}\propto 8=(10^{-1.1},10^{-1.5})$~(significantly noisier) levels.
We test the Kodak and CLIC Professional Validation datasets at full resolution with predetermined noise levels.
\section{More Experimental Results}
\label{sec2}
\begin{figure*}[t]
\centering
\includegraphics[width=0.49\linewidth, height=8cm]{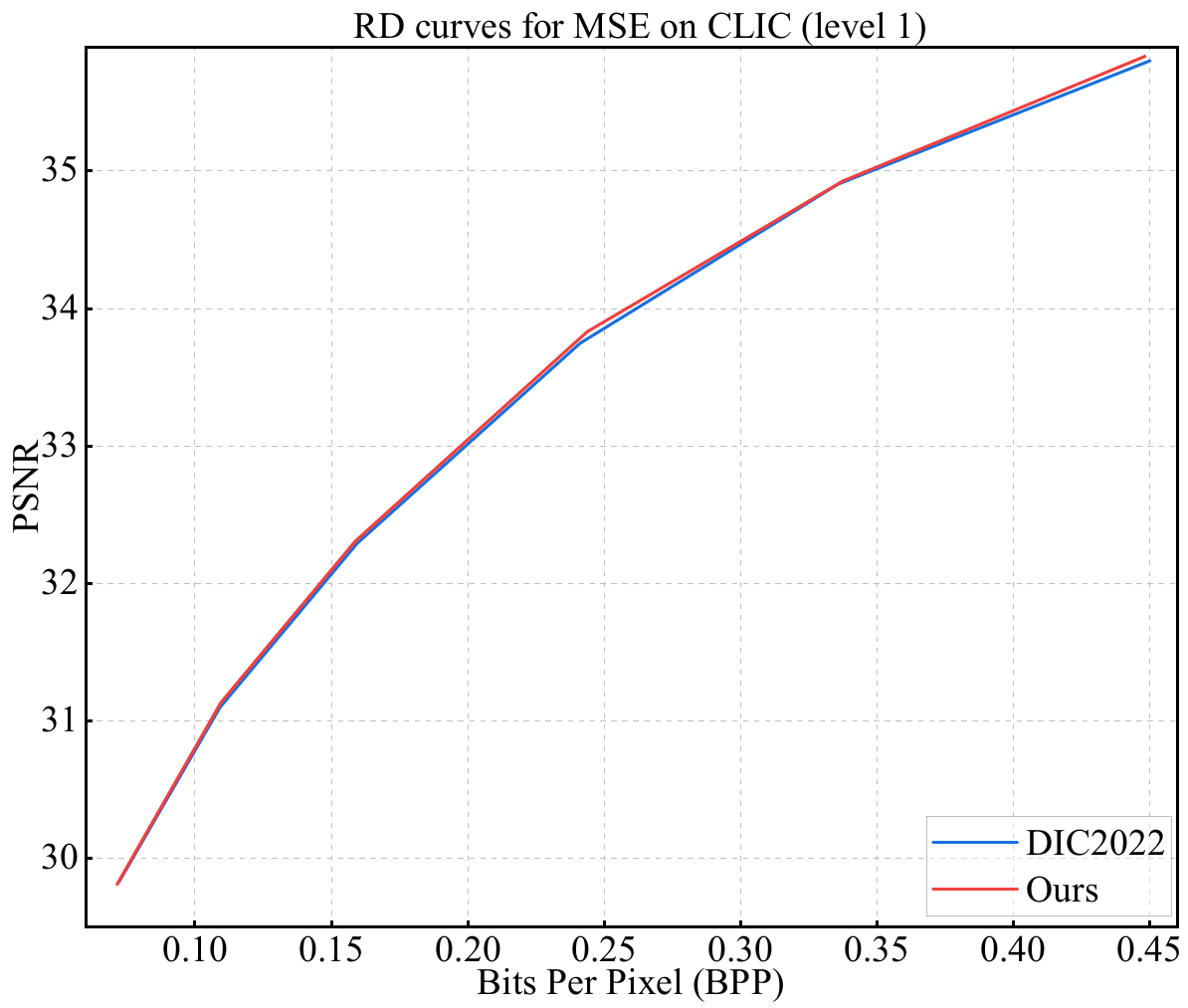}\hfill
\includegraphics[width=0.49\linewidth, height=8cm]{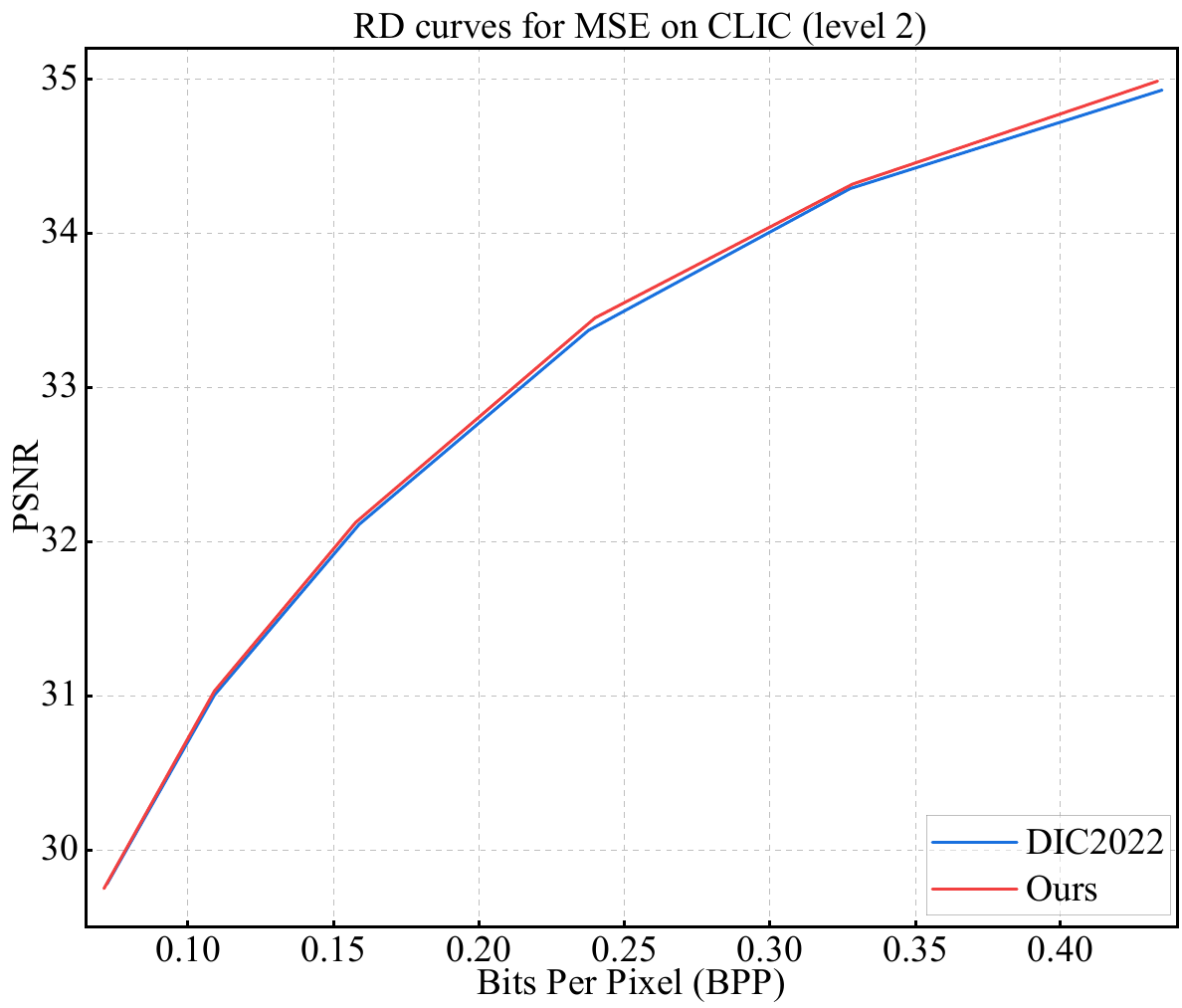}\\
\includegraphics[width=0.49\linewidth, height=8cm]{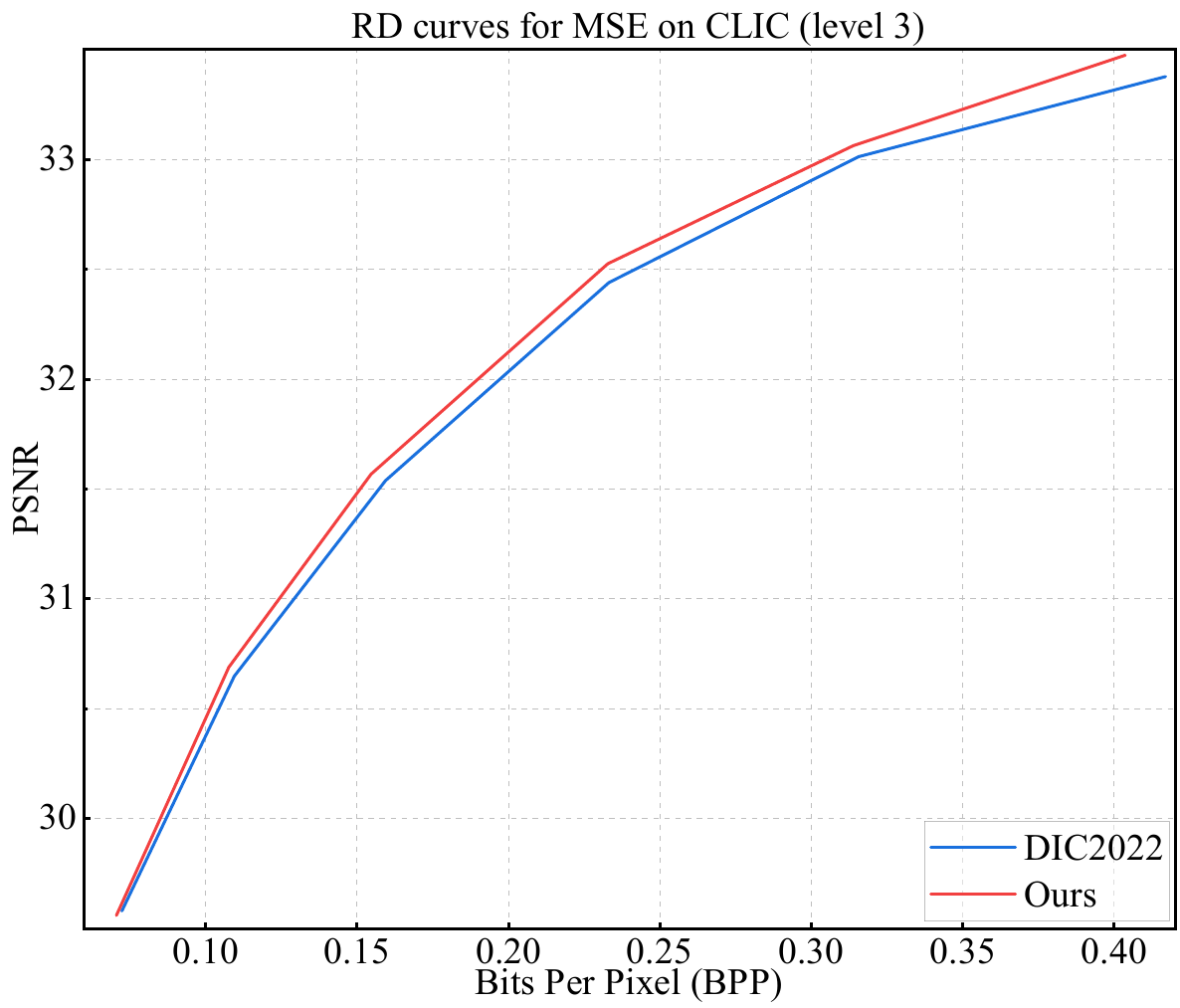}\hfill
\includegraphics[width=0.49\linewidth, height=8cm]{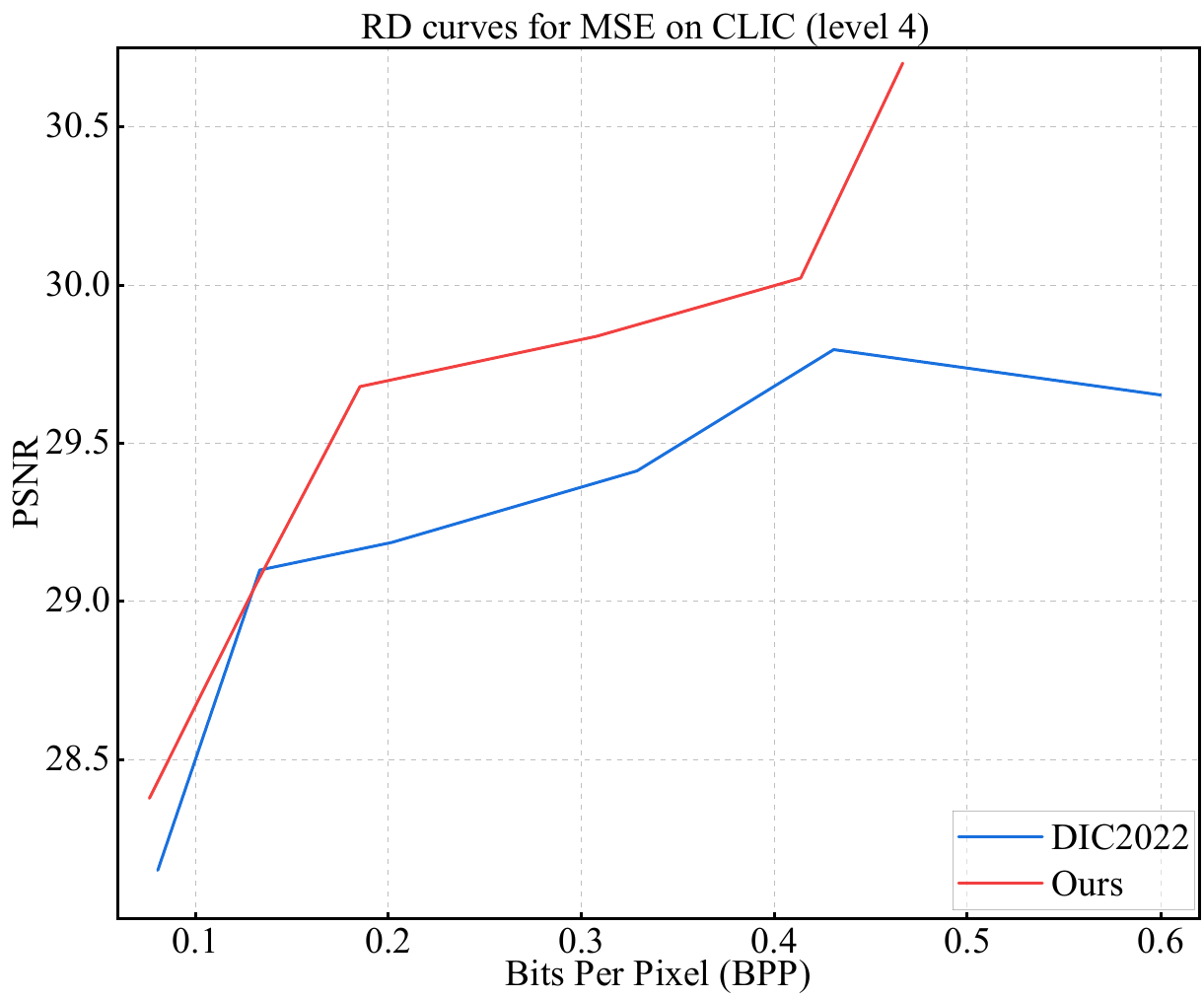}
\caption{RD curves on the CLIC dataset at various noise levels. Our method surpasses the state-of-the-art joint method, particularly at the high noise level. This denotes that the proposed SNR-aware branch efficiently captures valuable information through a combination of local and non-local features.}
\label{CLIC_individual}
\end{figure*}

Fig.~\ref{CLIC_individual} illustrates RD curves for MSE models on the CLIC dataset at individual noise levels to delve deeper into the impact of various noise levels. Our proposed joint method is better than the state-of-the-art joint method, especially at the high noise level. 
The signal-to-noise ratio~(SNR) in an image is usually low when the noise level is high. Therefore, it is difficult to achieve high-quality reconstructed images using only local features. With our proposed SNR-aware branch through local and non-local feature fusion, we can effectively be aware of different SNR regions in noisy images and obtain higher-quality reconstructed images.
\section{Compressing Generic Images}
\label{sec3}
We use the proposed network framework for compressing generic natural images. The training dataset is Flicker 2W~\cite{liu2020unified}. We discard images with a resolution of less than $512 \times 512$ in the dataset preprocessing stage. Randomly cropped patches with a resolution of $512 \times 512$ pixels are used to optimize the network. Our implementation relies on Pytorch~\cite{paszke2019pytorch} and an open-source CompressAI PyTorch library~\cite{begaint2020compressai}. The networks are optimized using the Adam~\cite{Kingma2015Adam} optimizer with a mini-batch size of 8 for approximately 2500000 iterations and trained on RTX 3090 GPUs.
The initial learning rate is set as $10^{-4}$ and decayed by a factor of 0.5 at iterations 800000, 1000000, 1200000, 1400000, 1600000, and 1800000. We train our models under 6 qualities, where $\lambda_{d}$ is selected from the set $\{0.0018, 0.0036, 0.0072, 0.013, 0.026, 0.0483\}$.
\par
\begin{figure*}[h]
\centering
\subcaptionbox{PSNR on Kodak}{\includegraphics[width = .49\linewidth, height=8cm]{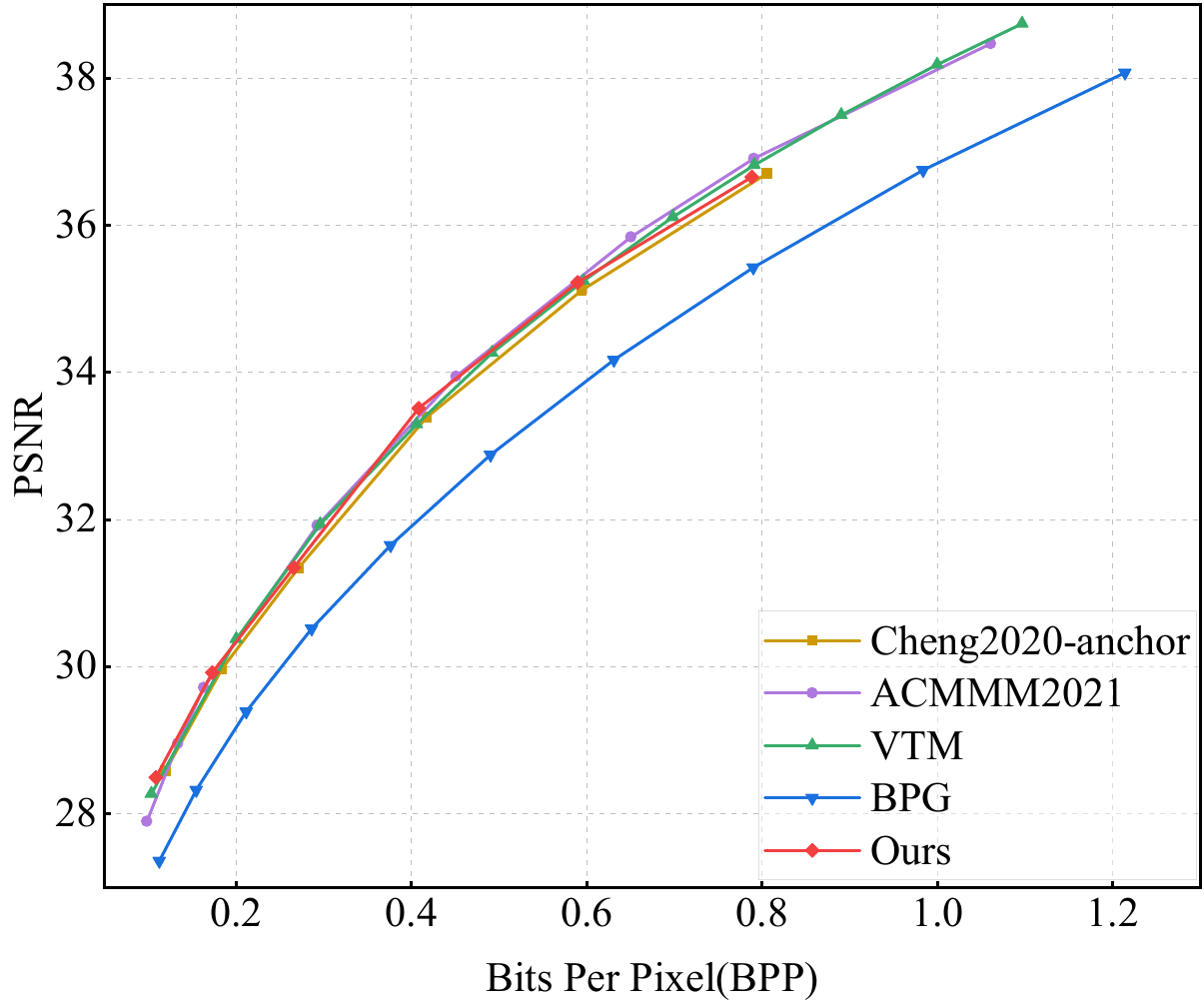}}\hfill
\subcaptionbox{PSNR on CLIC} {\includegraphics[width = .49\linewidth, height=8cm]{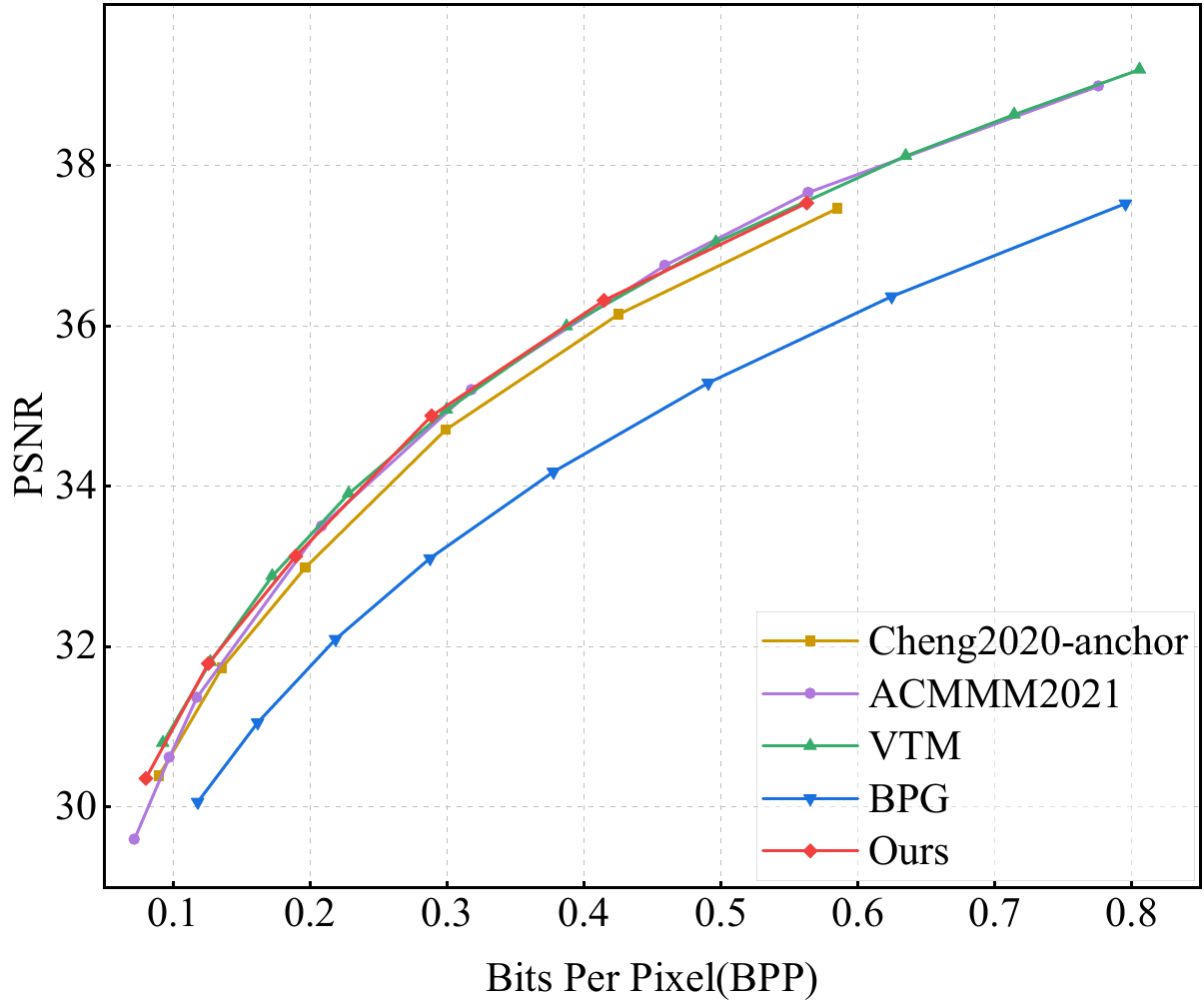}}\\[2ex]
\subcaptionbox{MS-SSIM on Kodak}{\includegraphics[width = .49\linewidth, height=8cm]{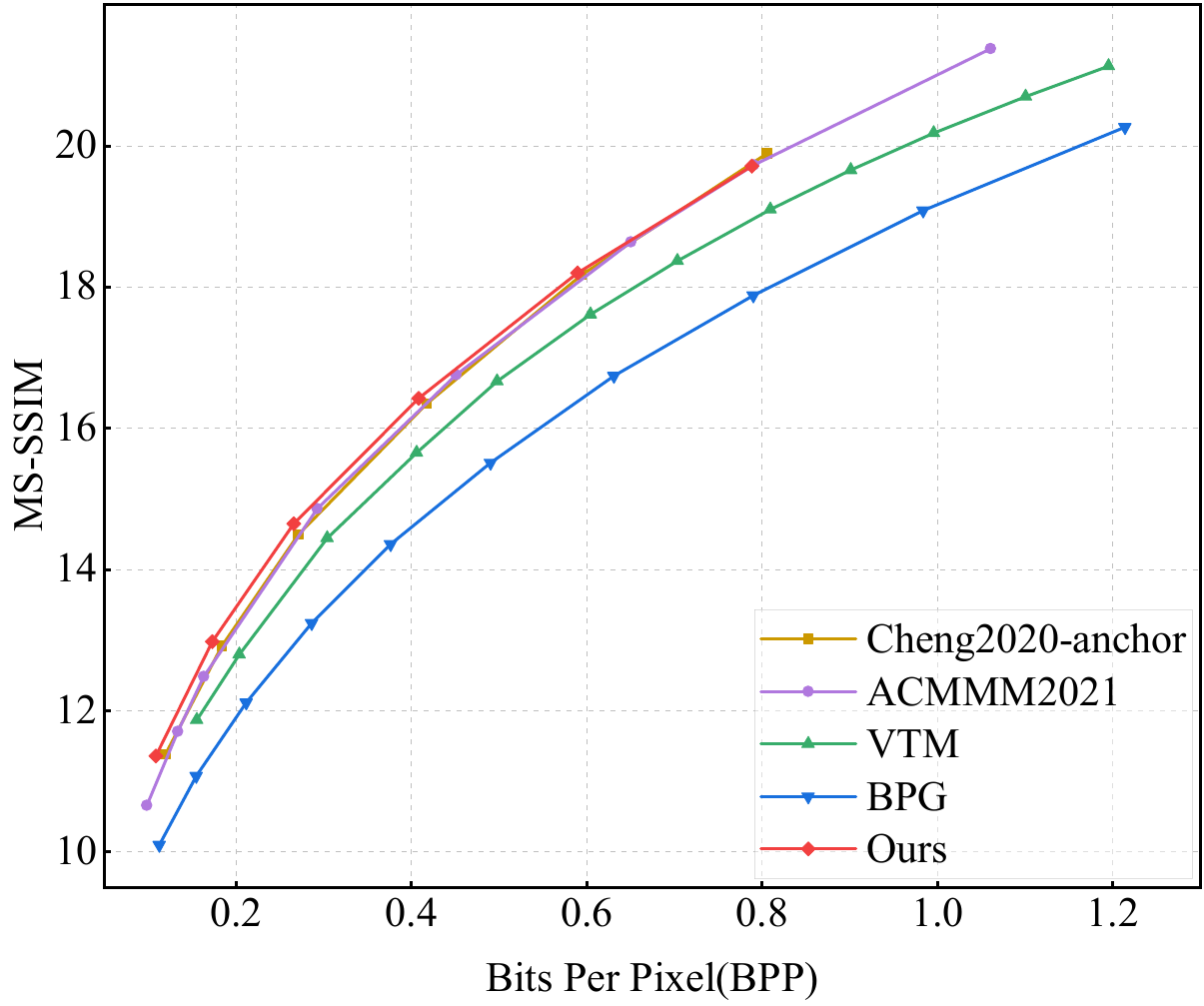}}\hfill
\subcaptionbox{MS-SSIM on CLIC}{\includegraphics[width = .49\linewidth, height=8cm]{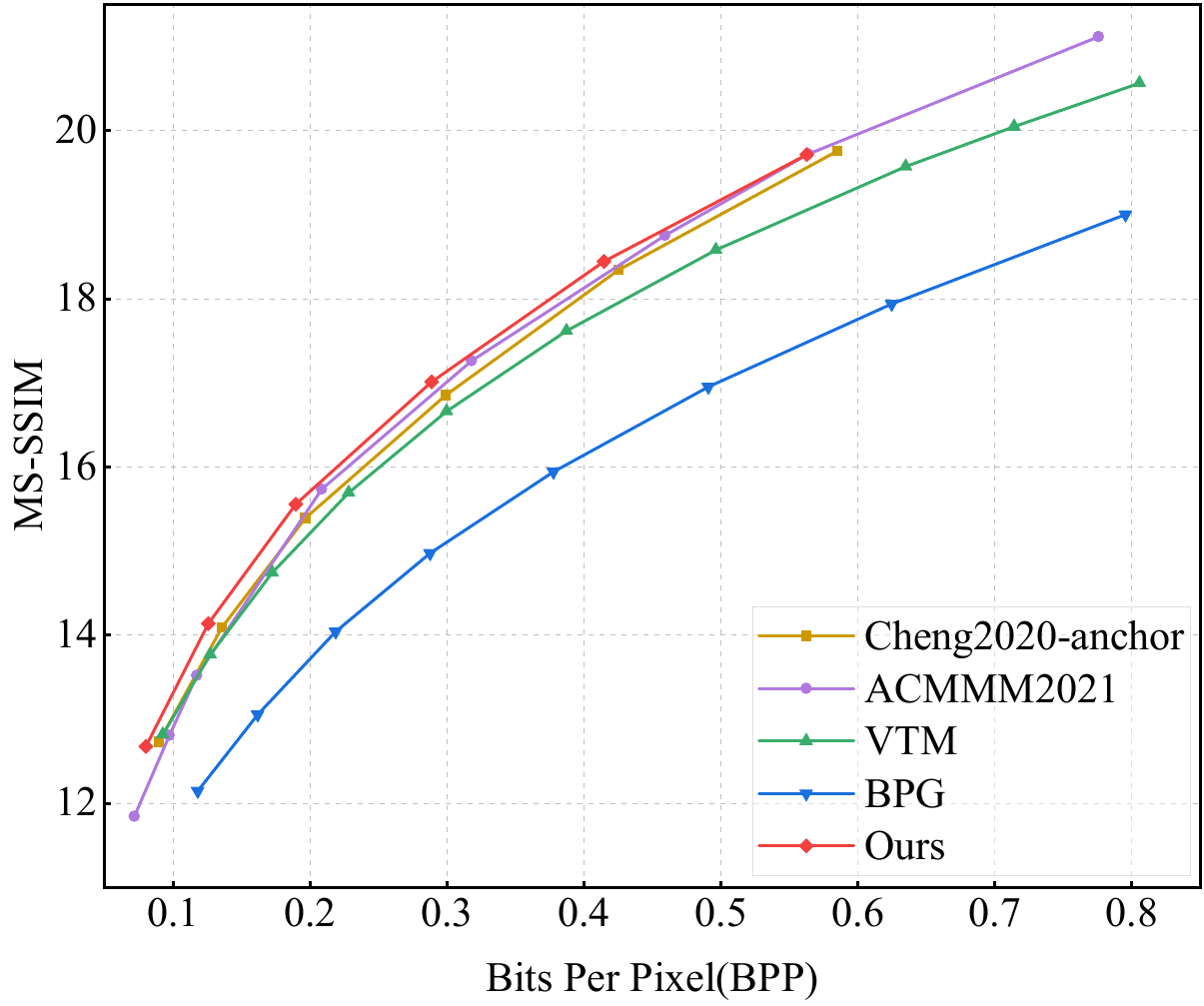}}
\caption{Rate-distortion performance curves aggregated over the Kodak and CLIC. MS-SSIM values converted to decibels $(-10log_{10}(1-\text{MS\mbox{-}SSIM}))$. (a)/(b) and (c)/(d) are results on Kodak and CLIC about PSNR and MS-SSIM, respectively. It indicates our proposed framework is also effective for compressing generic natural images.}
\label{RD_compression_curve}
\end{figure*}

We evaluate the rate-distortion performance on Kodak~\cite{kodak} and CLIC Professional Validation~\cite{CLIC2020} dataset. The rate-distortion performance comparison results of the proposed framework with the Cheng2020-anchor~\cite{cheng2020learned}, ACM MM2021~\cite{xie2021enhanced}, VTM~\cite{vvc} and BPG~\cite{bpg} are shown in Fig.~\ref{RD_compression_curve}. 
To achieve maximum compression performance, both VVC and BPG software are configured to utilize the YUV444 format.
The SNR-aware branch effectively captures local and no-local information about the original images which is useful for improving the rate-distortion performance. 
As the experimental results show, our proposed framework is better than the baseline~\cite{cheng2020learned} in terms of rate-distortion performance, indicating that our proposed joint framework is also effective for compressing generic natural images.
\section{Qualitative Results}
\label{sec4}
To further demonstrate the effectiveness of our method, we provide qualitative comparisons with the state-of-the-art joint method DIC2022~\cite{cheng2022optimizing}. Specifically, we show the visual results of the sample CLIC Professional Validation dataset~\cite{CLIC2020} images at noise level 4~($\text{Gain}\propto 8=(10^{-1.1},10^{-1.5})$).
These results show that our proposed SNR-aware joint method achieves superior reconstructed image quality even at lower bits per pixel~(BPP). In other words, our proposed joint method effectively leverages its awareness of different signal-to-noise ratio~(SNR) regions in the image, resulting in a significantly improved quality of the reconstructed image.

\begin{figure*}[]
\centering
\includegraphics[width=\linewidth,height=7.cm]{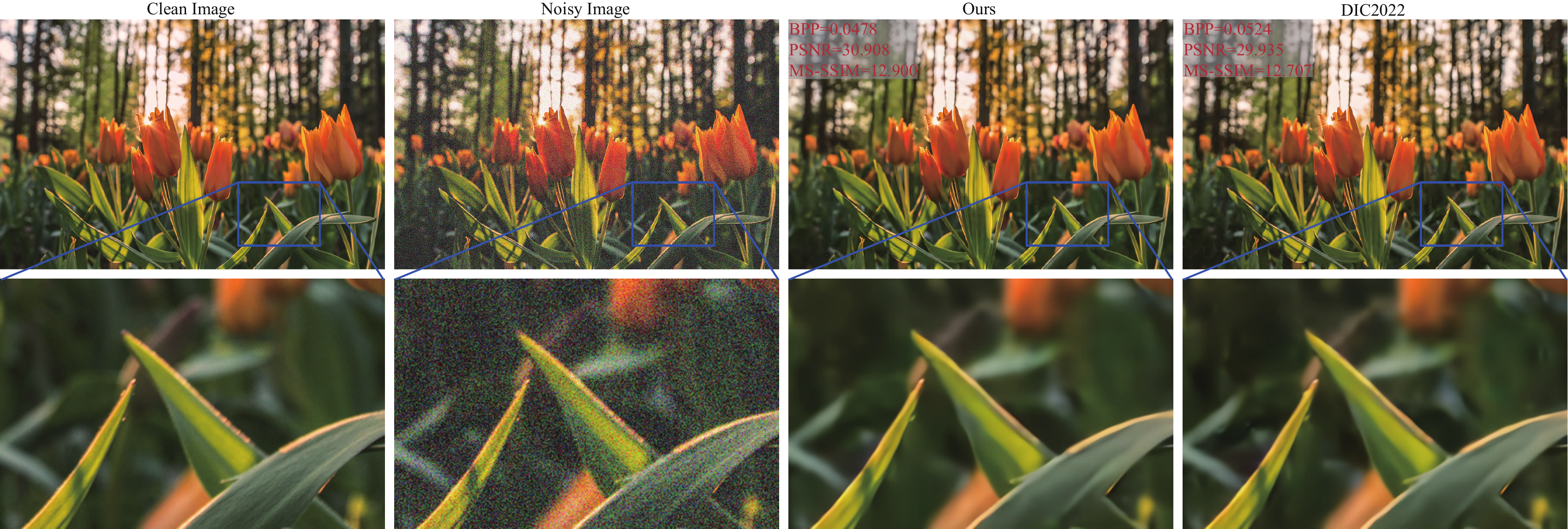}
\label{0002}
\end{figure*}

\begin{figure*}[]
\centering
\includegraphics[width=\linewidth,height=7.cm]{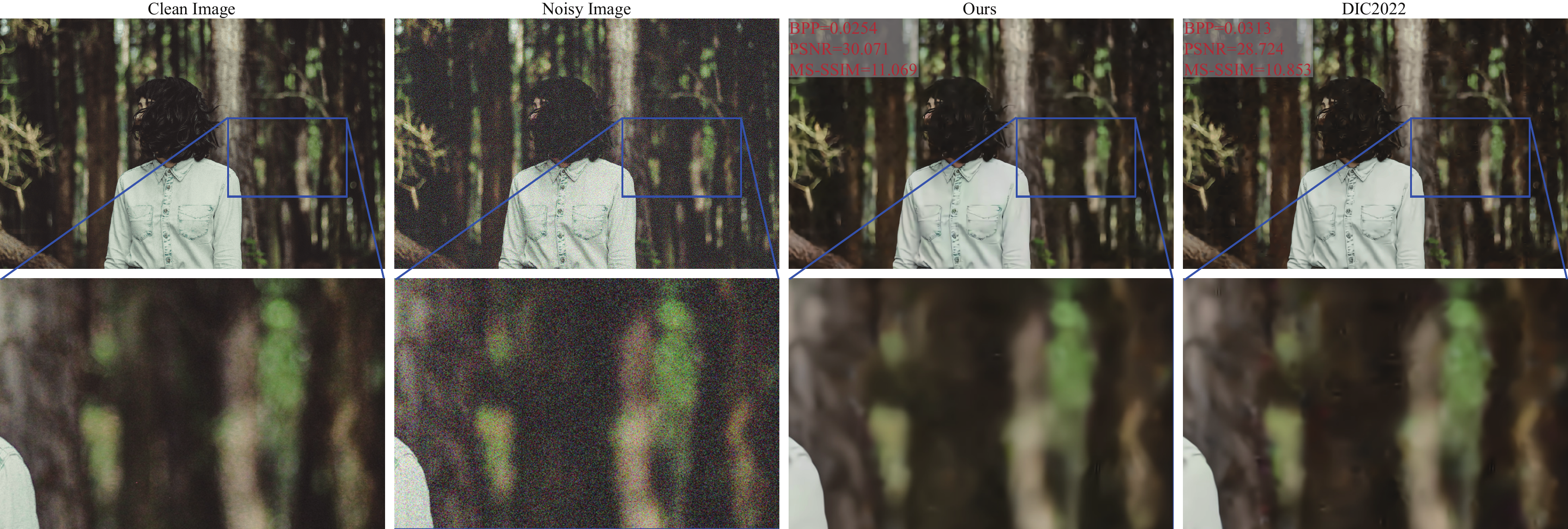}
\label{0004}
\end{figure*}

\begin{figure*}[]
\centering
\includegraphics[width=\linewidth,height=7.cm]{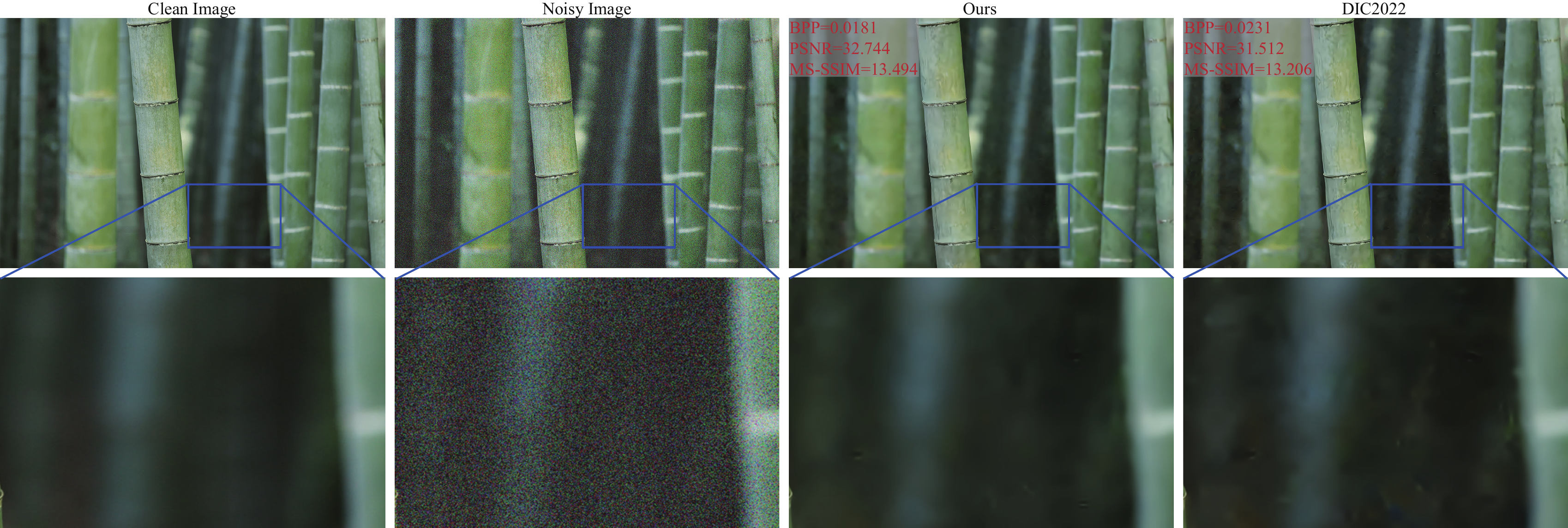}
\label{0019}
\end{figure*}

\begin{figure*}[]
\centering
\includegraphics[width=\linewidth,height=7.cm]{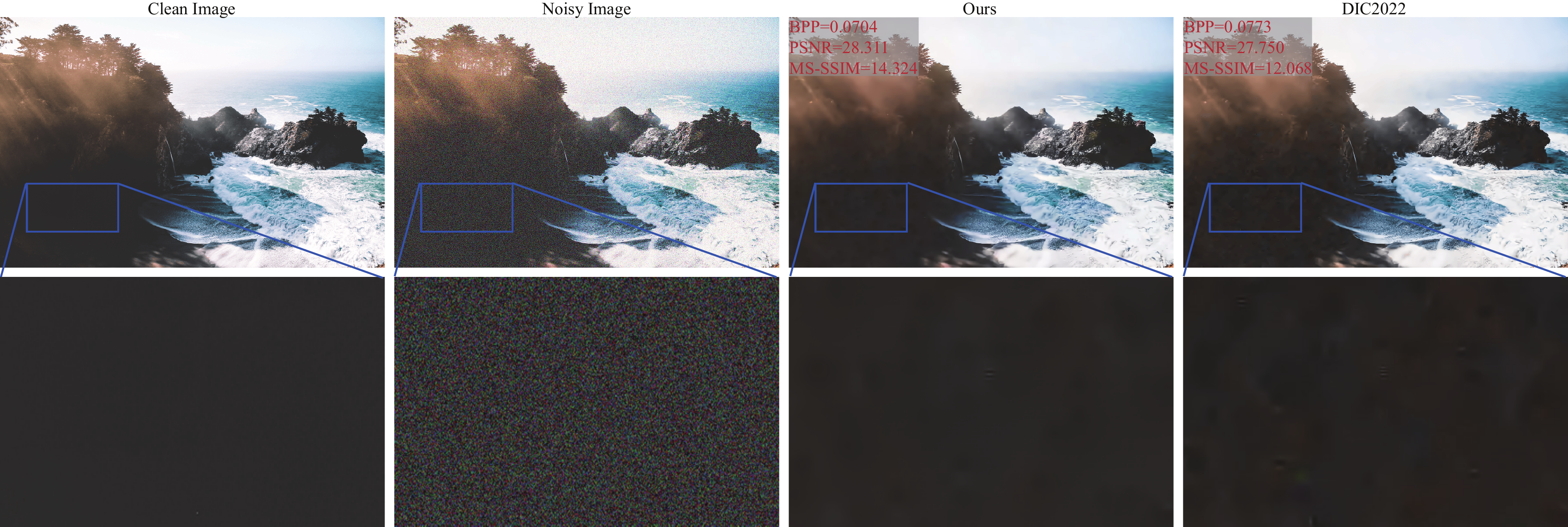}
\label{0023}
\end{figure*}

\begin{figure*}[]
\centering
\includegraphics[width=\linewidth,height=7.cm]{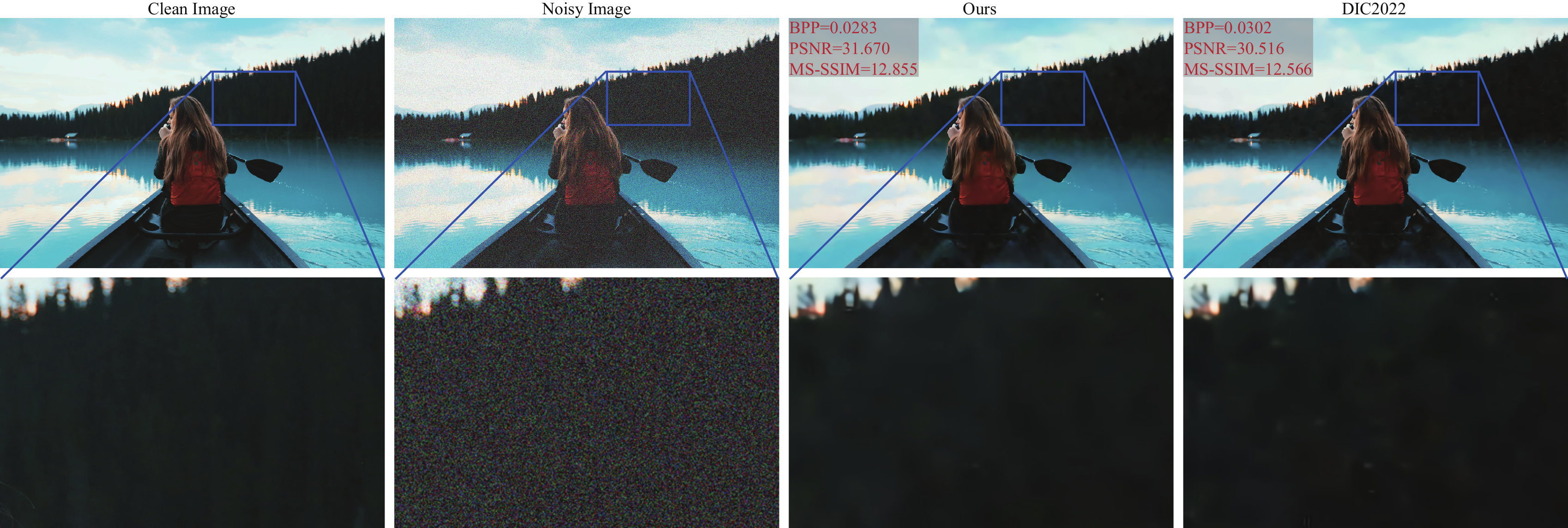}
\label{0029}
\end{figure*}

\begin{figure*}[]
\centering
\includegraphics[width=\linewidth,height=7.cm]{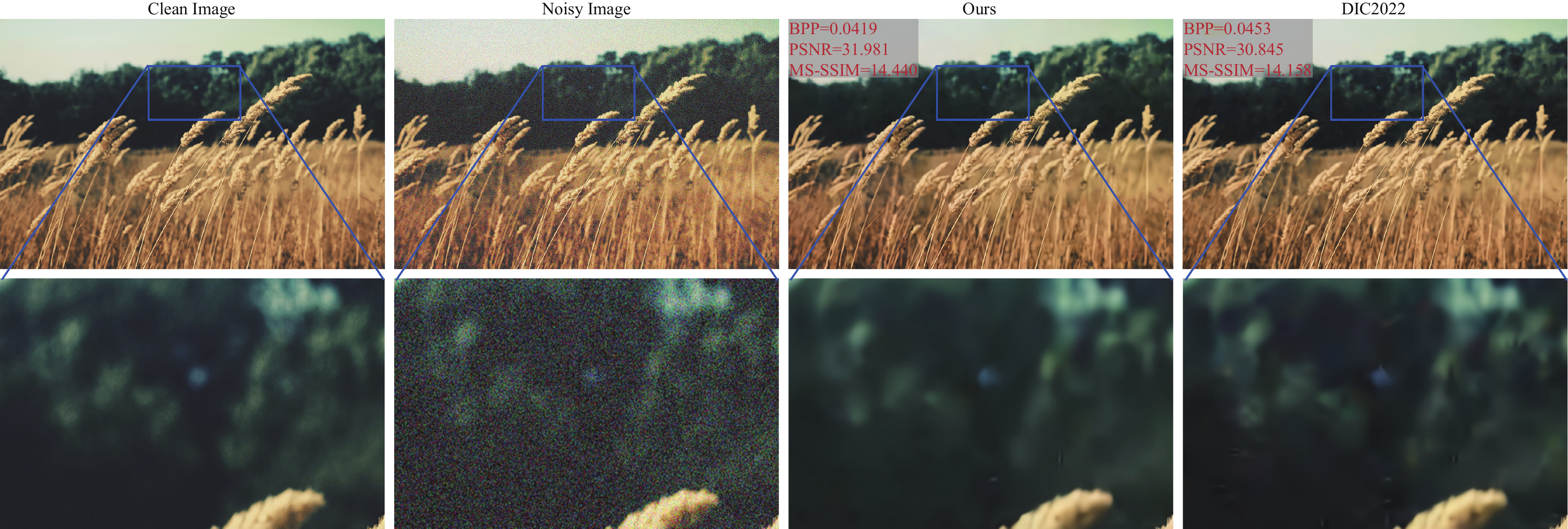}
\label{0040}
\end{figure*}


\end{document}